\documentclass[ aip, amsmath,amssymb, reprint, ]{revtex4-1}
\usepackage{float}
\usepackage{graphicx}
\usepackage{dcolumn}
\usepackage{bm}

\usepackage[T1]{fontenc}
\usepackage[english]{babel}
\usepackage{mathptmx}
\usepackage{xcolor}
\usepackage{hyperref}

\newcommand\crule[3][black]{\textcolor{#1}{\rule{#2}{#3}}} 

\newcommand{\be}{\begin{equation}}
\newcommand{\ee}{\end{equation}}
\newcommand{\bea}{\begin{eqnarray}}
\newcommand{\eea}{\end{eqnarray}}

\newcommand{\ec}[1]{\textcolor{black}{#1}}

\begin{document}


\title[]{Mechanical properties of Nucleic Acids and the non-local Twistable Wormlike Chain model}

\author{Midas Segers}
\author{Aderik Voorspoels}
\affiliation{Soft Matter and Biophysics Unit, KU Leuven, Celestijnenlaan
200D, 3001 Leuven, Belgium}
\author{Takahiro Sakaue}
\affiliation{Department of Physical Sciences, Aoyama Gakuin University, 5-10-1 Fuchinobe, Chuo-ku, Sagamihara, 252-5258, Kanagawa, Japan}
\author{Enrico Carlon}
\email{enrico.carlon@kuleuven.be}
\affiliation{Soft Matter and Biophysics Unit, KU Leuven, Celestijnenlaan
200D, 3001 Leuven, Belgium}

\date{\today}

\begin{abstract}
\ec{Mechanical properties of nucleic acids play an important role in many biological 
processes which often involve physical deformations of these molecules. At sufficiently 
long length scales (say above $\sim 20-30$ base pairs) the mechanics of DNA and RNA double 
helices is described by a homogeneous Twistable Wormlike Chain (TWLC), a semiflexible 
polymer model characterized by twist and bending stiffnesses. At shorter scales this 
model breaks down for two reasons: the elastic properties become sequence-dependent 
and the mechanical deformations at distal sites gets coupled. We discuss in this
paper the origin of the latter effect using the framework of a non-local Twistable Wormlike 
Chain (nlTWLC). We show, by comparing all-atom simulations data for DNA and
RNA double helices, that the non-local couplings are of very similar nature in 
these two molecules: couplings between distal sites are strong for tilt and twist 
degrees of freedom and weak for roll. We introduce and analyze a simple double-stranded 
polymer model which clarifies the origin of this universal distal couplings behavior.
In this model, referred to as the ladder model, a nlTWLC description emerges from the
coarsening of local (atomic) degrees of freedom into angular variables which describe the
twist and bending of the molecule. Differently from its local counterpart, the nlTWLC is 
characterized by a length-scale-dependent elasticity. Our analysis predicts that nucleic 
acids are mechanically softer at the scale of a few base pairs and are asymptotically 
stiffer at longer length scales, a behavior which matches experimental data.}
\end{abstract}

\maketitle

\section{Introduction}

\begin{figure}[b]
\includegraphics[width=\linewidth]{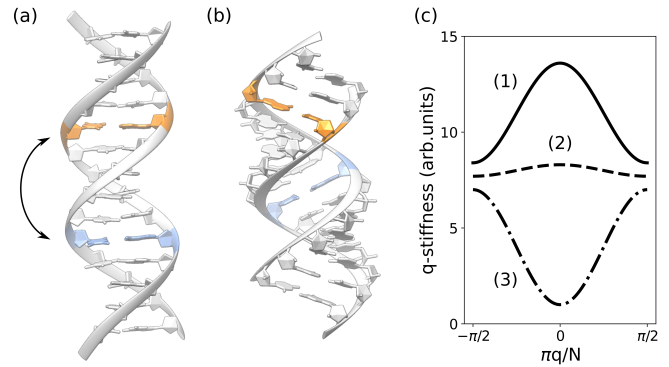}
\caption{Molecular structures of (a) DNA and (b) RNA double helices.
We focus here on the analysis of couplings between distal sites
along the helices, as those indicated in colors in (a) and (b). Such
interactions are characterized through the determination of momentum space
stiffnesses \cite{skor21}.  The plot (c) illustrates possible behaviors of
$q$-stiffnesses, with $q=0$ corresponding to the long wavelength limit:
(1) strong non-locality with stiffening at long distances (maximum at
$q=0$), (2) very weak non-locality and (3) strong non-locality with
softening at long distances (minimum at $q=0$).}
\label{fig:intro-DNA-RNA} 
\end{figure}

Mechanical properties of nucleic acids are of high relevance in several
biological processes (see e.g.\ Refs.~\onlinecite{agga20,dohn21,mari21}
for recent reviews) as these molecules are often deformed by
the action of ligands or by thermal fluctuations. A large number
of studies has provided a good deal of understanding of nucleic
acids mechanics, in particular via computer simulations of all-atom
\cite{lank03,noy12,pasi14,vela20,walt20} or of coarse-grained type
\cite{dans10,sulc12,skor17,chak18}.  While DNA is most commonly found
in a double helical form, RNA is usually a single stranded molecule,
but it can also form double helices in cells or viruses.  Due to the
different chemical nature of their constituent strands (deoxyribose
vs. ribose) DNA and RNA form helices with different geometries, known as
B and A-forms in the literature, see Fig.~\ref{fig:intro-DNA-RNA}(a,b).
For a recent review of the differences in the mechanics of DNA and
RNA see Ref.~\onlinecite{mari21}.  In this paper we are particularly
interested in characterizing non-local couplings which involve
non-consecutive base pairs \cite{skor21,sege21,lank09,noy12,esla11}
(Fig.~\ref{fig:intro-DNA-RNA}(a)).  The aim of this paper is to highlight
the similarities and differences between non-local interactions in DNA
and RNA and introduce a simple model which provides some quantitative
understanding on the general properties of non-local couplings of elastic
double-stranded molecules.

At the microscopic scale any conformation of a DNA or RNA double helix 
can always be given by the Cartesian coordinates of all the atoms. Such 
detailed description is however not very informative when one analyzes 
deformations involving several base pairs. For that purpose it is more 
convenient to use a set of coarse-grained coordinates describing for 
example bending or twist angles. It is via these coordinates that one
can define, for instance, bending and twist persistence lengths.
The number of coarse-grained coordinates used depends on the degree of
accuracy one wishes to achieve to describe local deformations. In the 
rigid base-pair model, base-pairs are modeled as rigid bodies \cite{lave10}.
The six variables tilt, roll, twist, shift, slide and rise describe the
rotations and displacements of consecutive base pairs. 
Tilt and roll describe the bending along the two possible bending directions 
of the helix. The twist describes the rotation along an axis perpendicular 
to the plane formed by a base pair. Shift, slide and rise describe the 
displacements of two consecutive base  pairs along the three axes. 
The even more detailed rigid base model includes six additional 
intra-basepair coordinates \cite{gonz13}. 

\ec{We will show here how, by passing from an atomistic description 
to a coarser one using angular coordinates (tilt, roll, twist), 
one naturally introduces non-local couplings in the system. 
In Sec.~\ref{sec:nonl-stiff} we
introduce the general formalism of the non-local Twistable Wormlike 
Chain (nlTWLC) which is our reference framework to describe coupling 
between distal sites in a homogeneous semiflexible twistable polymer. 
This model uses tilt, roll and twist as angular coordinates and 
neglects translational
degrees of freedom. We highlight the differences with the local 
TWLC and introduce the stiffness matrix in momentum space \cite{skor21},
which provides a very convenient formalism to discuss distal couplings.
In Sec.~\ref{sec:RNA} we analyze all-atom simulations data describing
elasticity of double stranded RNA and compare  with earlier DNA data
\cite{skor21}, within the nlTWLC framework. DNA and RNA data show strong
similarities which motivated us to look for a simple model that captures
the essential universal features of non-locality for both molecules. This
is the ladder model introduced in Sec.~\ref{sec:ladder}. This model
is sufficiently simple so that several quantities can be estimated
analytically and shows how non-distal coupling emerge from the coarsening
of ``atomistic'' degrees of freedom. The ladder model naturally explains
why couplings between distal sites are weak for roll and strong for tilt.
It also indicates that the twist-twist coupling between distal sites may
have a genuine non-local origin.  Sec.~\ref{sec:conclusion} concludes
the paper by highlighting the experimental relevance of our findings.  }

\section{The nlTWLC and stiffnesses in Fourier space}
\label{sec:nonl-stiff}

We consider here a description based on the parameters tilt ($\tau$),
roll ($\rho$) and twist ($\Omega$), using a rigid base pair model which
neglects translational stretching degrees of freedom. To describe the
deformations of the molecule we introduce the three dimensional vector
$\Delta_n$, with $n=0,1,\ldots N-1$ labeling the sites along the sequence,
defined as
\begin{equation}
\Delta_n \equiv \left( 
\tau_n - \langle \tau_n \rangle,
\rho_n - \langle \rho_n \rangle,
\Omega_n - \langle \Omega_n \rangle
\right)^\intercal    
\end{equation}
with $\langle . \rangle$ denoting equilibrium averages, so that $\langle
\Delta_n \rangle = 0$. Small deformations from equilibrium are usually
described within the harmonic approximation \cite{dohn21}, with an energy
given by
\begin{equation}
\beta E = \frac{a}{2} \sum_n \sum_m
{\Delta}^\intercal_n M^{(n)}_m {\Delta}_{n+m},
\label{eq:energy_real}
\end{equation}
with $\beta=1/k_BT$ the inverse temperature, $k_B$ the Boltzmann
constant and $a$ the average distance between consecutive base pairs
($a=0.34$~nm for DNA, whilst $a=0.37$~nm for dsRNA).  It is well-known
that local mechanical properties are very strongly influenced by the
sequence composition, our aim however is to characterize the properties
of non-local couplings averaged over several different sequences.
Therefore we neglected sequence inhomogeneities, as such in the rest
of the paper we will drop the label $(n)$ from $M^{(n)}_m$.  The energy
\eqref{eq:energy_real} includes possible couplings between distal sites
$n$ and  $n+m$, which are encoded in the $3 \times 3$ matrix $M_m$.
Setting to zero all matrices $M_m=0$ for $m \geq 1$, corresponds to
considering a local model. In this limit Eq.~\eqref{eq:energy_real}
describes the usual discrete Twistable Wormlike Chain (TWLC). We refer to
the model given by \eqref{eq:energy_real} with generic distal couplings,
as the non-local Twistable Wormlike Chain (nlTWLC).  Several all-atom
\cite{lank03} and coarse-grained simulations \cite{skor17} indicated
that in DNA variables on distal sites are correlated, which implies
a coupling between them. In previous papers \cite{skor21,sege21} we
have investigated non-local couplings in DNA, which we briefly review
(Sec.~\ref{sec:nonl-stiff}), before analyzing these interactions in RNA
(Sec.~\ref{sec:RNA}) and discussing a simple model of non-local elasticity
(Sec.~\ref{sec:ladder}).

A full account of the elastic behavior of a polymer model with non-local 
interactions has been presented in Ref.~\onlinecite{skor21}. Here we
give a brief summary of the main results. Non-local couplings are more 
conveniently described in a momentum-space representation via discrete 
Fourier transforms of the deformation parameters \cite{skor21}
\begin{equation}
    \widetilde \Delta_q = \sum_{n=0}^{N-1}  \Delta_n\, e^{-2\pi i q n/N}
    \label{def:Deltaq}
\end{equation}
where $q = -(N-1)/2, -(N-3)/2, \ldots ,(N-3)/2, (N-1)/2$ (for $N$ odd). 
As the vector $\Delta_n$ contains real numbers, one finds by complex 
conjugation $\widetilde \Delta_q^* = \widetilde \Delta_{-q}$.
The nlTWLC \eqref{eq:energy_real} in $q$-space then takes the form 
\begin{equation}
\beta E = \frac{a}{2N} \sum_q 
{\widetilde\Delta}^\dagger_{q} \widetilde{M}_q {\widetilde\Delta}_q,
\label{eq:energy_fourier}
\end{equation}
where the matrices $\widetilde{M}_q$ and $M_n$ are related to one another via 
Fourier transform \cite{skor21}
\begin{equation}
    \widetilde{M}_q  = \sum_m  M_m \, e^{-2\pi i q m/N}
    \label{eq:mtilde}
\end{equation}
Formally, the step from \eqref{eq:energy_real} to \eqref{eq:energy_fourier}
requires a system with periodic boundary conditions, which ensure translational
invariance. However, the difference between a linear or circular molecule at
small deformations is only linked to boundary terms, which do not influence
the large $N$ bulk behavior.

Note that in the local model limit (TWLC), corresponding to $M_m =
0$ for $m \geq 1$, the $q$-space stiffness matrix $\widetilde{M}_q$
becomes $q$-independent.  A weak dependence indicates that the stiffness
is predominantly determined by local interactions. Conversely, a strong
$q$-dependence reflects strong non-local effects.  Non-local couplings
give rise to length-scale-dependent elasticity \cite{skor21}, i.e.\ the
molecule can become either stiffer or softer when its elastic behavior
is probed at longer length scales. A maximum in the q-stiffness at $q=0$
indicates then a stiffening at increasing lengths, as the  $q \to 0$
limit corresponds to the asymptotic long length scale limit $N \to
\infty$ in real space.  Fig.~\ref{fig:intro-DNA-RNA}(c) shows examples
of $q$-space stiffnesses for molecules which are stiffer (1) or softer
(3) at increasing lengths.

\section{Non-local couplings: DNA vs. RNA}
\label{sec:RNA}

In order to assess the presence of possible off-site couplings in RNA
duplexes, all-atom MD simulations were performed for several 20-mer
sequences (for simulation details see Appendix \ref{appendix:all_atom}).
Tilt, roll and twist were extracted from all-atom data using the
Curves+ software \cite{lave09}. The vectors $\widetilde{\Delta}_q$
were obtained from the discrete Fourier transform \eqref{def:Deltaq}
and the $q$-stiffness matrix was calculated from the inversion of the
covariance matrix \cite{skor21}
\begin{equation}
    \left\langle\Tilde{\Delta}_q\Tilde{\Delta}_q^\dagger\right\rangle
    =\frac{N}{a}\widetilde{M}_q^{-1}
    \label{eq:covariance}
\end{equation}
The stiffness matrix in $q$-space takes the form
\begin{equation}
    \widetilde{M}_q=
    \begin{pmatrix}
        \widetilde{A}^t_q && \widetilde{D}_q && \widetilde{B}_q\\
        \widetilde{D}_{-q} && \widetilde{A}^r_q && \widetilde{G}_q\\
        \widetilde{B}_{-q} && \widetilde{G}_{-q} && \widetilde{C}_q
    \end{pmatrix}
    \label{eq:q_space_matrix_3d}
\end{equation}
where the (real) diagonal elements $\widetilde{A}^t_q$,
$\widetilde{A}^r_q$ and $\widetilde{C}_q$ are the tilt, roll and
twist stiffness. The matrix $\widetilde{M}_q$ is hermitian hence
the transposed off-diagonal elements are related by complex
conjugation: $\widetilde{D}_q^*=\widetilde{D}_{-q}$,
$\widetilde{B}_q^*=\widetilde{B}_{-q}$ and
$\widetilde{G}_q^*=\widetilde{G}_{-q}$. Each of these elements have in
principle a real and an imaginary part which are respectively symmetric
and anti-symmetric in $q$, as follows from elementary properties of
Fourier transforms.

\begin{figure}[t]
\includegraphics[width=\linewidth]{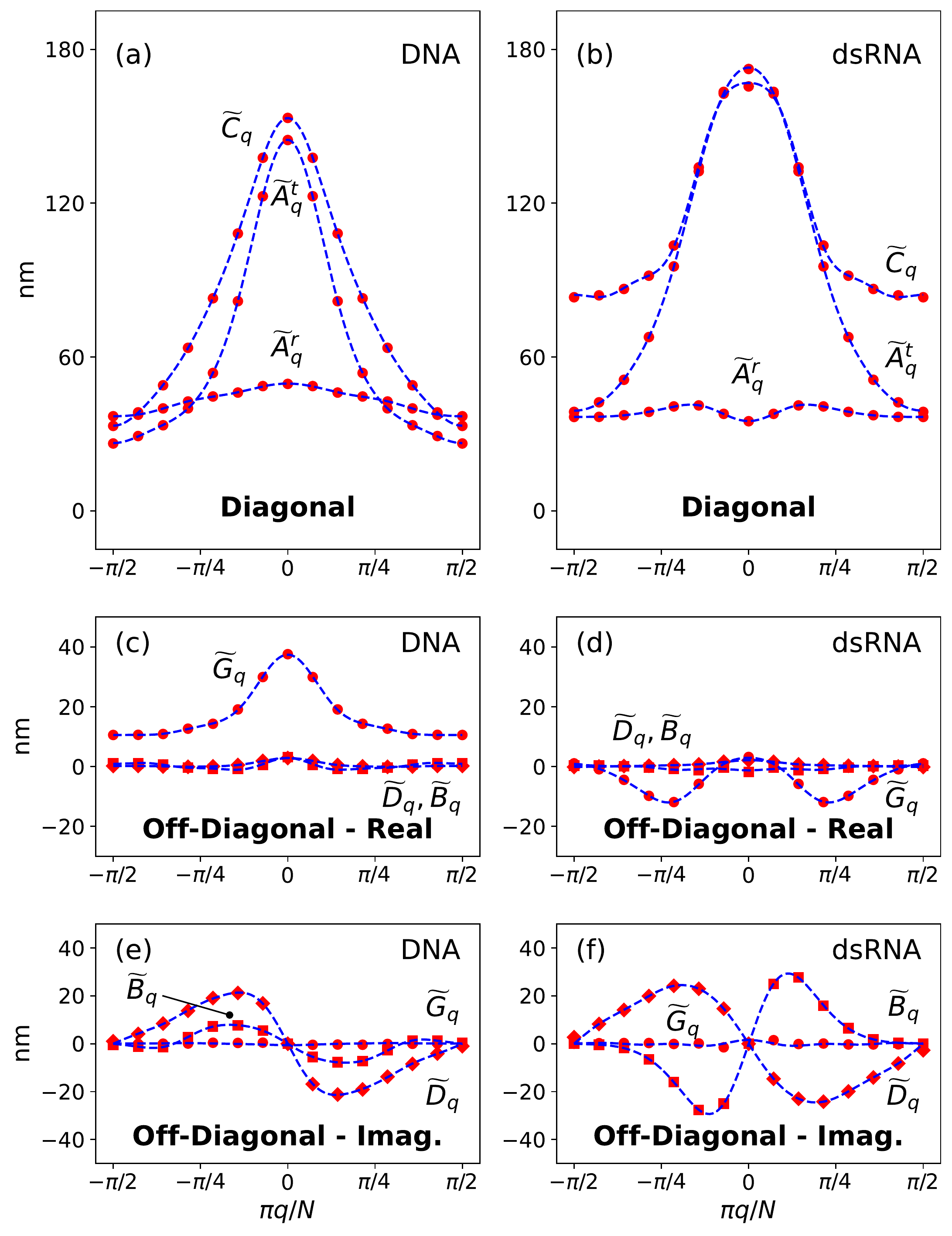}   
\caption{Summary of comparison between elements of momentum space
stiffness matrices $\Tilde{M}_q$ for DNA (left, Ref.~\onlinecite{skor21})
and RNA (right, this work) as obtained from all-atom simulations. The
three panel show diagonal elements (a,b), and the real (c,d) and imaginary
(e,f) of the off-diagonal coupling terms. The red circles are calculated
directly from \eqref{eq:covariance}, the blue dashed line represents a
fit truncating the Fourier series \eqref{eq:mtilde} to a finite number
of terms. The fitting parameters are given in Appendix \ref{appB}.  }
\label{fig:DNA-RNA}
\end{figure}

Figure~\ref{fig:DNA-RNA}(a,b) compares the diagonal terms of DNA (left,
Ref.~\onlinecite{skor21}) and RNA (right) stiffnesses. Both nucleic
acids show a strong $q$-dependence for the tilt-tilt $\widetilde{A}^t_q$
and twist-twist $\widetilde{C}_q$ couplings, indicating the presence of
significant off-site interactions for these degrees of freedom. Roll
deformations, instead, couple predominantly locally for both DNA and
RNA, as seen from the weak dependence on $q$ of the roll-roll coupling
$\widetilde{A}^r_q$.  Off-diagonal stiffnessess for DNA and RNA are
shown in Fig.~\ref{fig:DNA-RNA}(c,d,e,f).

\begin{figure}[t]
\includegraphics[width=1.0\linewidth]{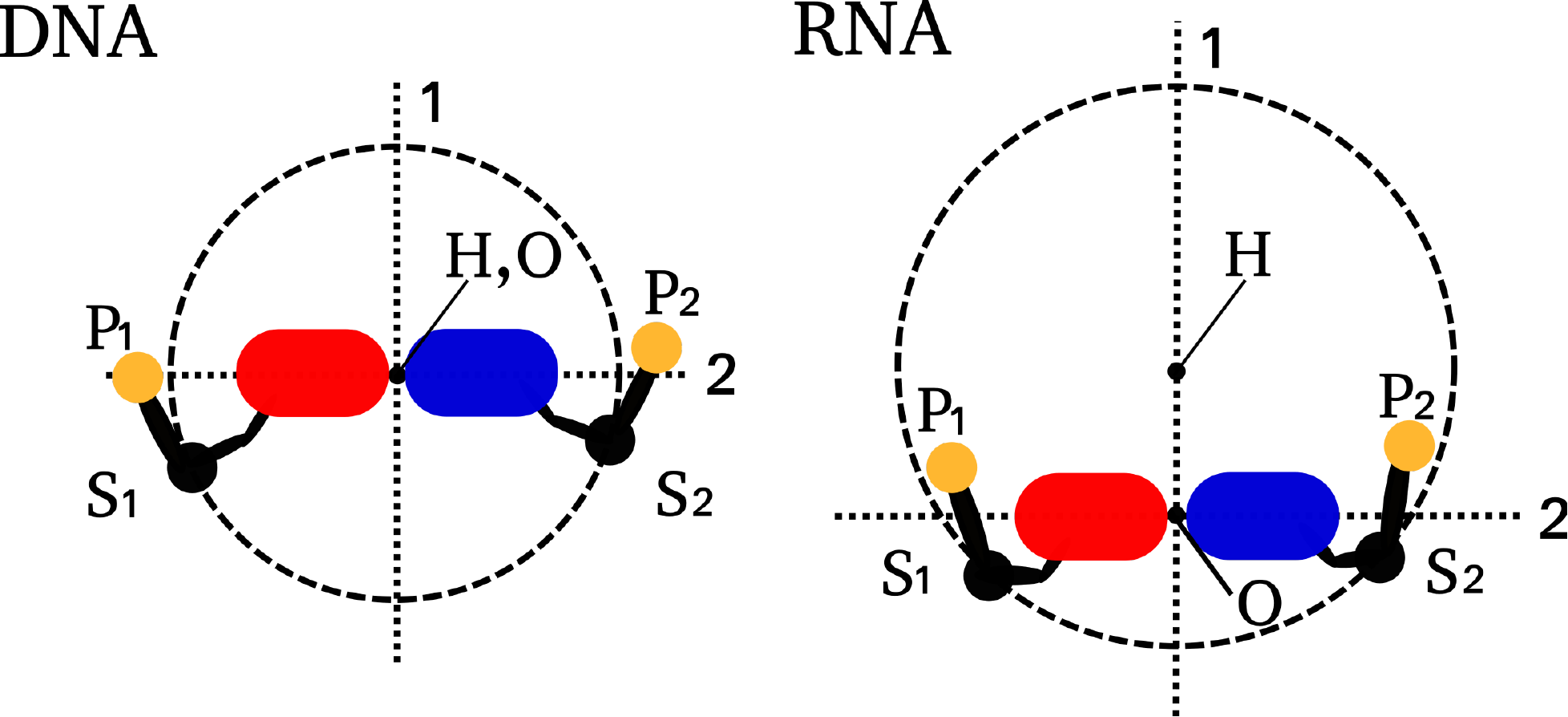}
\caption{Top view of the molecular structure of DNA and
RNA (we draw the RNA structure as planar, in reality the
base pair plane is not perpendicular to the helical axis, see
Fig.~\ref{fig:intro-DNA-RNA}(b)). The determination of the tilt, roll and
twist parameters is performed by assigning an orthonormal triad associated
to each base pair.  Triads are centered in ``O'', the midpoint of the
basepairs, ``H'' denotes the position of the helical axis.  For both
molecules the triad is formed by the axis 1 and axis 2 and by a third
axis orthogonal to both. In DNA 1 is a symmetry axis, but not 2, which
is the origin of the twist-roll coupling \cite{mark95}. The molecular
structure of RNA suggests a very weak twist-roll coupling as 2 is close
to a symmetry axis (see text).}
\label{fig:tbc}
\end{figure}

A striking difference between DNA and RNA is that the former
has a large twist-roll coupling ($\text{Re} \, \widetilde{G}_q$,
Fig.~\ref{fig:DNA-RNA}(c,d)). \ec{Such a coupling } was predicted
from the general symmetry of the DNA molecule \cite{mark94} \ec{and}
has several consequences on the structure, fluctuations and dynamics of
DNA \cite{skor18,cara19,nomi19,fosa21}, but it is much weaker in RNA. To
gain some insights about this difference one can consider the structure
of DNA and RNA molecules.  \ec{For this purpose a top view on the plane
perpendicular to the base pairs is shown in } Fig.~\ref{fig:tbc}. \ec{Here
}S$_1$ and S$_2$ identify the positions of the \ec{sugars} of the two
backbones, while 1 and 2 \ec{label} the two axes used in the calculations
of the deformation parameters.  \ec{Rotations about axis 1 from one
base pair to the next correspond to tilt deformations, while rotating
about axis 2 produces roll deformations}. In both DNA and RNA the axis
1 is close to a symmetry axis: a 180$^\circ$ rotation around this axis
interchanges the positions of the two strands.  \ec{On the other hand,
the symmetry about axis 2 is broken by the major/minor groove differences
in DNA, which is the origin of a twist-roll coupling \cite{mark95}.  RNA
however, from analysis of the molecular structure, seems to have a weaker
asymmetry on axis 2 than DNA, which is likely the reason for the weaker
twist-roll coupling. Of course there cannot be exact symmetries due to the
chemical differences of the bases.  Nonetheless,} we note that $\text{Re}
( \widetilde{G}_q) $, although weak, is larger than other off-diagonal
terms in RNA, see Fig.~\ref{fig:DNA-RNA}(d). Fig.~\ref{fig:DNA-RNA}(e,f)
compares the imaginary parts of the off-diagonal terms, which are
antisymmetric in $q$.  These terms vanish at $q=0$ and have a small
influence on the overall elastic properties of the molecules.

\begin{figure}[t]
\includegraphics[width=\linewidth]{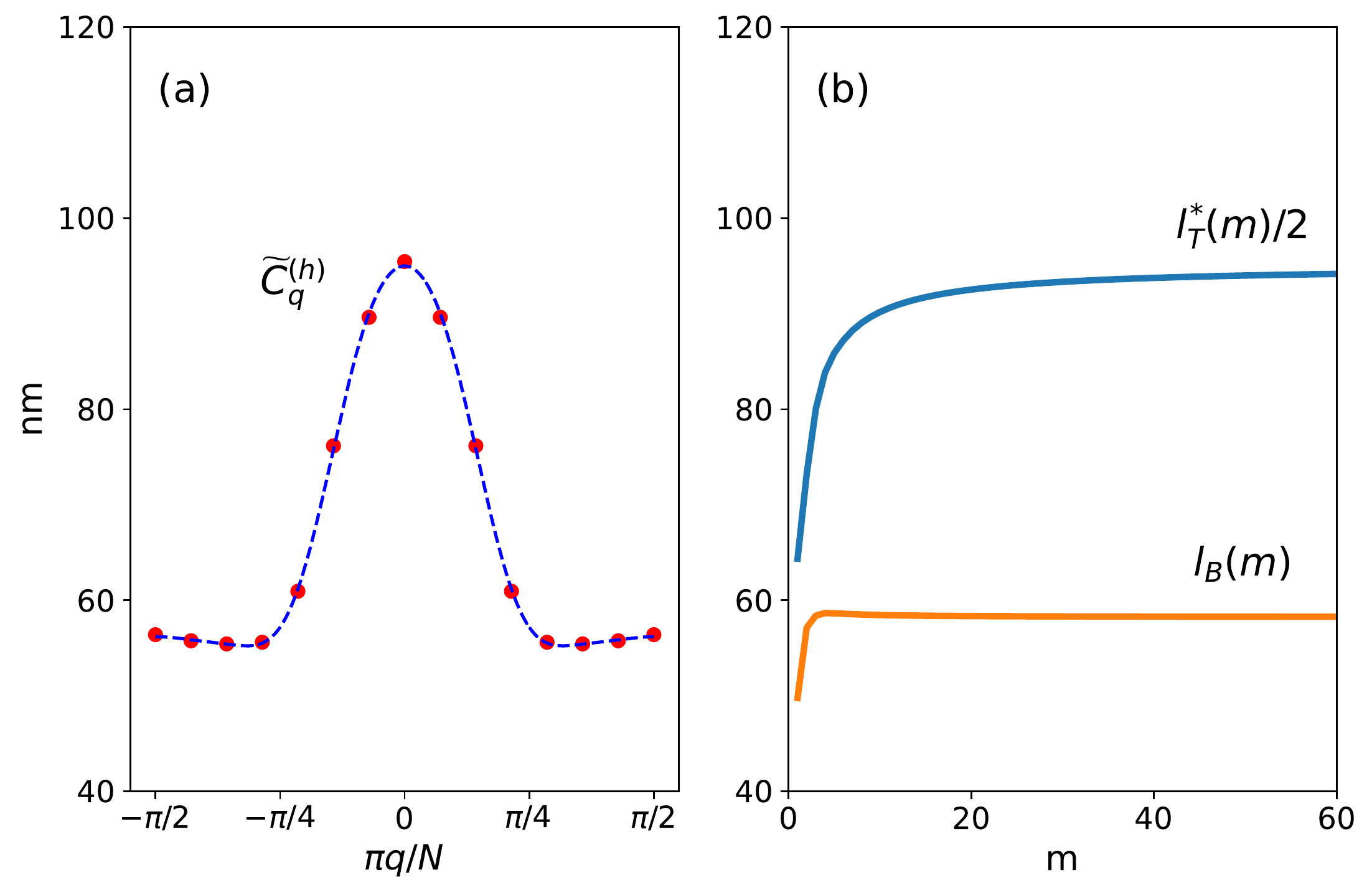} 
\caption{(a) Q-stiffnesses of the helical twist of RNA as
obtained from all-atom data. (b) Length-scale-dependence of the
twist (blue) and bending (orange) persistence lengths for RNA
duplexes. The length-dependence of $l^*_T(m)/2$ is derived from
equation \eqref{RNA:lT} using the $q$-stiffnesses displayed in
(a).  $l_B(m)$ is acquired using equation \eqref{def:lBm} along
with the attained momentum-space stiffnesses exhibited in Figure
\ref{fig:DNA-RNA}(b).}
\label{fig:Persistence_lengths}
\end{figure}

The $q$-dependence of the stiffness parameters implies
length-scale-dependent persistence lengths \cite{skor21}. A twistable
polymer is usually characterized by two persistence lengths associated
to bending and twist deformations. We discuss the twist persistence
length first. The $\widetilde{C}_q$ of Fig.~\ref{fig:DNA-RNA}(b)
is the stiffness for a twist around an axis perpendicular to the
base-pair plane, which is tilted with respect to the helical axis,
see Fig.~\ref{fig:intro-DNA-RNA}(b). To describe the experimentally
relevant twist around the helical axis, the variable helical twist,
$\Omega_n^{(h)}$, as well as the average helical rise, $a^{(h)}$,
are used \cite{lave09}. The helical twist correlation function, for a
stretch of $m+1$ base pairs, is defined as
\begin{equation}
    \left<\cos\left(a^{(h)}\sum_{n=0}^{m-1}\Omega_n^{(h)}\right)\right>=e^{-ma^{(h)}/l^*_T(m)}
    \label{eq:twist_cor}
\end{equation}
and its decay is governed by the twist correlation length
$l^*_T(m)$. Due to couplings between distal sites $l^*_T(m)$ becomes
length-scale-dependent. To compute it one can use the relation
\cite{skor21}
\begin{eqnarray}
\frac{1}{l^*_\text{T}(m)}
&=& \frac{a^{(h)}}{2\pi m} \int_{-\pi/2}^{\pi/2} \frac{\sin^2 my}{\sin^2 y}
\frac{\langle |\widetilde{\Omega}_q^{(h)}|^2 \rangle}{N} \, dy,
\label{RNA:lT}
\end{eqnarray}
valid for an infinitely long polymer chain $N \to \infty$, where $y
\equiv \pi q/N$ and where $\widetilde{\Omega}_q^{(h)}$ indicates the
discrete Fourier transform of $\Omega_n^{(h)}$.  For a $q$-independent
stiffness and from the elementary integral
\begin{eqnarray}
    \int_{-\pi/2}^{\pi/2} \frac{\sin^2 my}{\sin^2 y} dy = m \pi
\end{eqnarray}
one finds a length-scale-independent $l^*_T$, as for the ordinary TWLC.
In general, for large $m$ one gets from \eqref{RNA:lT} the following
asymptotic expansion \cite{sege21}
\begin{equation}
    l_T^*(m) = l_T^{*}(\infty) \left( 1 - \frac{B}{B+m} \right) + \ldots
\end{equation}
with B a coefficient and where the dots indicate exponentially
small terms.  We obtained the helical twist from simulation
data using the Curves+ software \cite{lave09}, given in
Fig.~\ref{fig:Persistence_lengths}(a). The blue dashed line is
an interpolation of the all-atom data using a truncated Fourier
series as explained in Appendix \ref{appB}.  These data were
used to compute numerically the integral in Eq.~\eqref{RNA:lT}
and obtained the torsional persistence length $l_T^*(m)$, shown in
Fig.~\ref{fig:Persistence_lengths}(b).

To estimate the bending persistence length we used the standard tilt
and roll data. Curves+ does not seem to provide these deformations with
respect of the helical axis, therefore we will compute a tangent-tangent
correlator
\begin{equation}
    \langle \hat{t}_n \cdot \hat{t}_{n+m} \rangle = e^{-ma/l_B(m)}
    \label{eq:tan-tan_cor}
\end{equation}
with $\hat{t}_n$ perpendicular to the base plane and thus tilted with
respect to the helical axis. In the limit of small intrinsic roll and
twist, the following expression can be used to estimate the bending
persistence length
\begin{equation}
    \frac{1}{l_\text{B}(m)} = \frac{a}{2\pi m} 
    \int_{-\pi/2}^{\pi/2} \frac{\sin^2 my }{\sin^2 y} \,
    \frac{\langle|\widetilde\tau_q|^2 \rangle + 
    \langle |\widetilde\rho_q|^2 \rangle}{N} \, dy
    \label{def:lBm}
\end{equation}
which was computed numerically from the interpolated all-atom data
(dashed lines of Fig.~\ref{fig:DNA-RNA}(b,d,f)). The bending persistence
length thus obtained is shown in Fig.~\ref{fig:Persistence_lengths}(b). We
note that while $l_B$ shows a very modest length-scale-dependence, this
dependence is much stronger for $l_T^*$, in agreement with what has
been previously observed for DNA \cite{skor21}. Both nucleic acids are
torsionally softer at short scales and stiffer at longer length scales.

Table~\ref{tab:RNA_data} compares the bending and torsional persistence
lengths as obtained in this work from $q$-stiffnesses data, via
Eqs.~\eqref{RNA:lT} and \eqref{def:lBm}, with other simulation
and experimental data from the literature. The bending persistence
length reported here is in line with reports from both experimental and
simulation studies.  Much less has been reported on the twist persistence
length, especially from the experimental side. Nonetheless the values
calculated here based on momentum space considerations are in agreement
with the existing literature.

\begin{table}[t]
    \centering
\begin{tabular}{c@{\hskip 0.2in}c@{\hskip 0.2in}c@{\hskip 0.2in}c}
    \hline
    \hline
    $l_B$~(nm)  & $l_T^{*}/2$~(nm)  & Method & Ref.\\
    \hline
     {$\ec{58}\pm1$} & {$\ec{94}\pm2$} & all-atom MD     & this work \\
     {$\ec{59}\pm1$} & {$\ec{100}\pm2$} & all-atom MD$^*$ & this work \\
     $59 \pm 2$ & $100 \pm 2$  & Magnetic tweezers & [\onlinecite{lipf14}]\\
     $60 \pm 3$ & $-$  &  Optical tweezers & [\onlinecite{herr13}]\\
     $66\pm1$ & $75 \pm 6$ & all-atom MD & [\onlinecite{mari17}] \\
     $63.8\pm0.7$ & $-$  & Magnetic tweezers/ AFM & [\onlinecite{Abels2005}]\\
     $78.9\pm3.4$ & $108 \pm 3 $ & all-atom MD &  [\onlinecite{Chen20}] \\
    \hline
    \hline
\end{tabular}
\caption{Summary of comparisons of RNA duplex persistence lengths as
obtained from the analysis of all-atom data in this paper and experimental
and simulation data from existing literature. The error on the values
reported in this paper was estimated from the standard deviation of
the data across different sequences.  \ec{We provide two estimates:
one is obtained by using the average over all simulated sequences and
another (marked with '*') removing three outlier sequences showing a
stiffness behavior strongly deviating from the average, see Appendix
\ref{appendix:all_atom}.} The reported values correspond well to the
experiments \cite{lipf14}.  It should however be noted that even though
the error is small questions remain regarding the proper definition of
$l_{B}$ when comparing to experiments.}
\label{tab:RNA_data}
\end{table}

\section{Origin of non-local elasticity: insights from simple models}
\label{sec:ladder}

Having discussed non-local couplings in RNA and underlined the
similarities with earlier DNA data \cite{skor21}, we now investigate the
origin of these couplings using a minimal model of elastic polymer chain
with twist and bending degrees of freedom. The model, with an appropriate
choice of couplings, can reproduce qualitatively the $q$-stiffness
spectrum which characterizes the elastic behavior of nucleic acids.

\subsection{Ladder Model}

\begin{figure}[t]
\includegraphics[width=\linewidth]{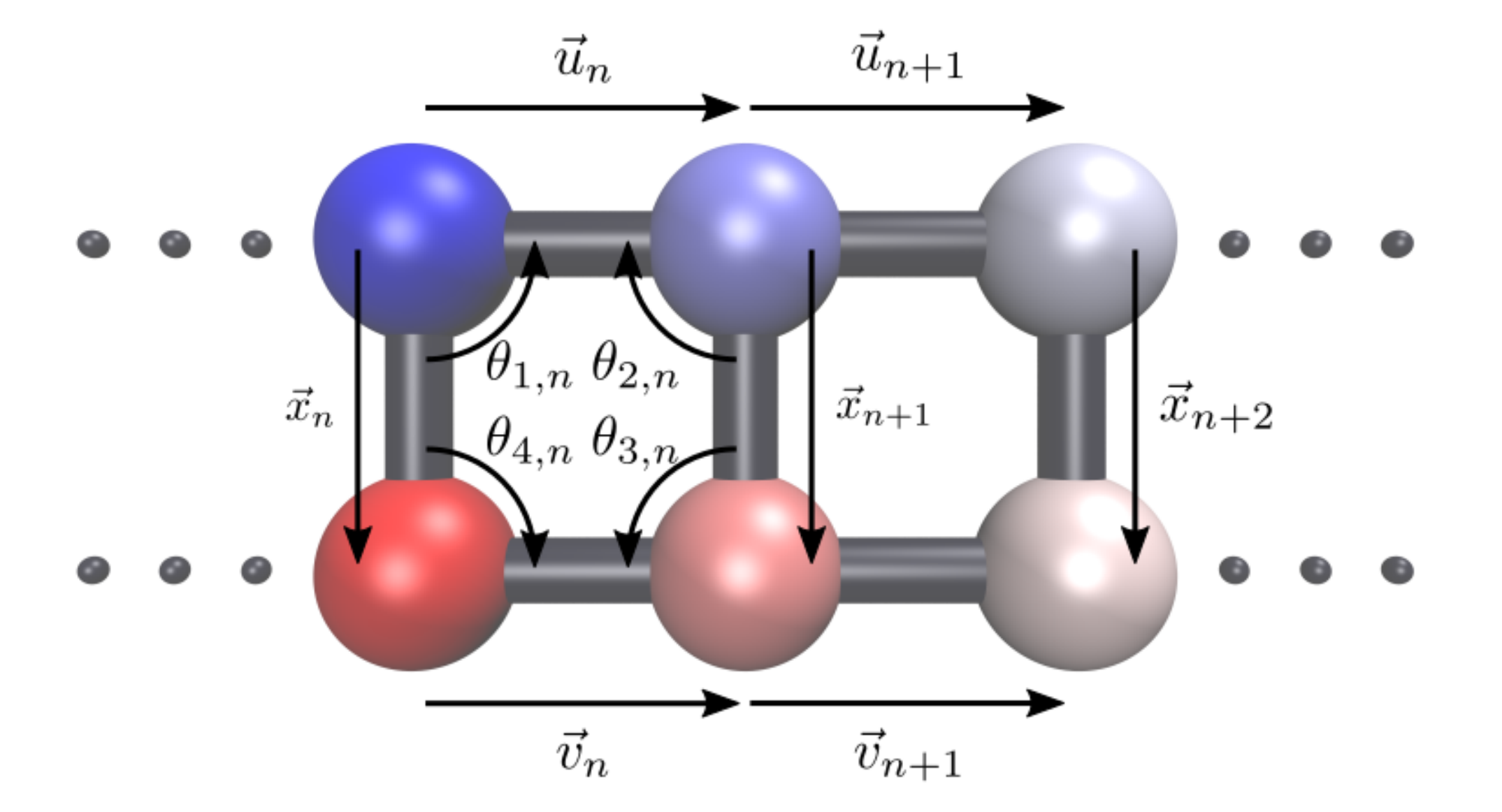} 
\caption{Schematic of the ladder model in its ground-state configuration. 
Intra- and inter-strand bond vectors are denoted with $\vec{u}$, $\vec{v}$ 
and $\vec{x}$ respectively. }
\label{fig:ladder_flat}
\end{figure}

We consider an homogeneous mechanical system which consists of a set
of point-masses organised into two distinct strands, as illustrated
in Fig.~\ref{fig:ladder_flat}.  We refer to it as the ladder
model. Neighbouring masses interact with one another through four
distinct interaction types: bonds, backbone rigidity, angular rigidity
and stacking, each characterized by a corresponding stiffness. To
describe a conformation we use vectors $\vec{u}_n$ and $\vec{v}_n$
which connect neighboring masses along the two strands and vectors
$\vec{x}_n$ to describe the orientations of the rungs of the ladder
(Fig.~\ref{fig:ladder_flat}). We indicate with $\theta_{i,n}$ the
angles formed between the backbones and rung vectors, as shown in
Fig.~\ref{fig:ladder_flat}.  The energy of the model is given by
\begin{eqnarray}
        \beta E &=& \frac{K_{s}}{2} \sum_n \left[(|\Vec{x}_n|-l_0)^2+
         (|\Vec{u}_n|-l_0)^2+(|\Vec{v}_n|-l_0)^2 \right] \nonumber\\
         &-& K_{wlc}\sum_n \left(\hat{u}_n \cdot \hat{u}_{n+1} +
         \hat{v}_n \cdot \hat{v}_{n+1} \right) \nonumber\\
         &-&
        K_{a}\sum_n \sum_{i=1}^4\sin\theta_{i,n} -
        K_{st}\sum_n \hat{x}_n \cdot \hat{x}_{n+1}
    \label{eq:total_energy_ladder_model}
\end{eqnarray}
where $\beta=1/k_BT$ and where we used $\hat{u}_n \equiv \vec{u}_n/|
\vec{u}_n|$, $\hat{v}_n \equiv \vec{v}_n/| \vec{v}_n|$ and $\hat{x}_n
\equiv \vec{x}_n/| \vec{x}_n|$ to denote unit vectors.  $K_s$ is the
spring constant which governs the stretching energy of the bond vectors
and $l_0$ the rest length. Each backbone is modelled as a Worm-Like-Chain
(WLC) such that the alignment of subsequent bond vectors i.e. $\vec{u}_n$,
$\vec{u}_{n+1}$ and $\vec{v}_n$, $\vec{v}_{n+1}$ is energetically
favorable. The interaction strength is indicated as $K_{wlc}$.  The model
also contains an angular coupling, favoring $\theta_{i,n}=\pi/2$ angles
between adjacent backbone and rung bonds (the angles $\theta_{i,n}$ are
defined in Fig.~\ref{fig:ladder_flat}). The angular stiffness is governed
by the parameter $K_a$.  Additionally the model has a stacking-type
of interaction, with stiffness $K_{st}$, favoring the alignment of
consecutive rung vectors $\vec{x}_n$ and $\vec{x}_{n+1}$.

The lowest energy state of the system consists of a flat ladder with
bond vectors of fixed length $|\Vec{x}_n| = |\Vec{u}_n| = |\Vec{v}_n| =
l_0$, with the backbone vectors $\Vec{u}_n$, $\Vec{v}_n$ all parallel and
orthogonal to the rung vectors $\Vec{x}_n$. Thermal excitations distort
the ladder inducing bending and twist deformations whose magnitude
depends on the values of the stiffness constants $K_s$, $K_{wlc}$,
$K_a$ and $K_{st}$. We have set the stiffnesses such that at the level
of neighboring base pairs the distortion of angles and bond lengths is
weak, e.g.
\begin{equation}
    \left| \theta_{i,n} - \frac{\pi}{2} \right| \ll 1, \quad
    \left| \left| \Vec{x}_n \right| - l_0 \right| \ll l_0
\end{equation}

\subsubsection{Elementary excitations}

\begin{figure}[t]
\includegraphics[width=0.47\textwidth]{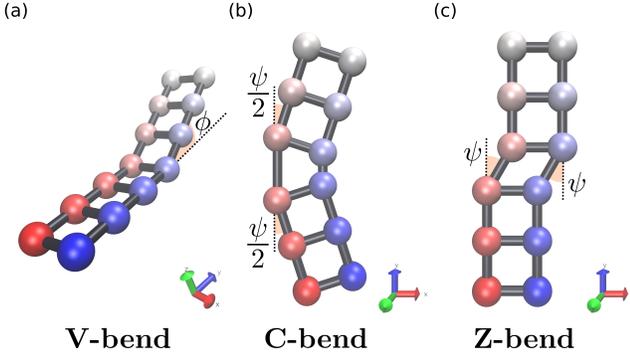}
\caption{Elementary bending deformations of the ladder model. (a)
Example of roll deformation, a bending out of the ladder plane. (b,c)
Examples of tilt deformations, which are bendings within the
ladder plane. Whereas the bending (a) - here referred to as
``V-bend'' - is localized to a single ladder rung, (b) and (c) -
denoted as ``C-bend'' and ``Z-bend'' respectively - consist of the
tilting of two consecutive rungs. Whilst the two bending angles
in a C-bend are identical, they are opposite in the Z-bend. The
excitation energies corresponding to each deformation are given in
Eqs.~\eqref{def:DEVrho}, \eqref{def:DECtau} and \eqref{def:DEZtau}.}
\label{fig:ladder_deformations}
\end{figure}

Before considering the elastic behavior of the ladder model and the
$q$-stiffnesses, we analyze three elementary bending deformations which
are useful for the following discussion.  The calculation gives some
indications on the general behavior of the system.  These deformations are
shown in Fig.~\ref{fig:ladder_deformations}.  The case (a) is referred
to as a roll deformation, while (b) and (c) are tilt deformations,
in analogy with the nomenclature used for nucleic acids \cite{lave09}.
We refer to (a) as a V-bend, to (b) as a C-bend and to (c) as a Z-bend.
Now, we will estimate the energy cost associated to the three bends.

The V-bend (a) is a rotation of part of the ladder around a
rung. Bonds are not stretched, the inner angles maintain their
ground state value $\theta_{i,n}=\pi/2$ and rungs remain parallel.
From \eqref{eq:total_energy_ladder_model}, we obtain the following energy
for this deformation
\begin{equation}
   \beta \Delta E_V^\rho = 2 K_{wlc} (1-\cos \phi) 
    \label{def:DEVrho}
\end{equation}
where $\phi$ is the rotation angle.
This is analogous to a bending deformation in a simple ``local'' WLC
model, the factor $2$ accounting for the fact that the two backbones
are bent.  The C-bend tilt of Fig~\ref{fig:ladder_deformations}(b)
keeps the rungs of the ladder unstretched, but there is some stretching
in the backbone bonds.  Minimizing this stretching energy we get from
\eqref{eq:total_energy_ladder_model} the following excitation energy
for the C-bend deformation
\begin{eqnarray}
    \beta\Delta E_C^\tau &=& K_s l_0^2 \sin^2 \frac{\psi}{2} + 
    4 \left(K_{wlc} + K_a \right) \left(1-\cos \frac{\psi}{2} \right) 
    \nonumber \\
    &&+ K_{st} (1-\cos \psi)
    \label{def:DECtau}
\end{eqnarray}
The C-bend describes a sequence of two bends of an angle $\psi/2$
each.  All the four microscopic stiffnesses $K_s$, $K_{wlc}$, $K_a$ and
$K_{st}$ contribute to this excitation energy.  Finally, the Z-bend tilt
deformation of type (c) does not involve bond stretching or stacking
interaction, as it is only influenced by backbone bending and angle
deformation. We find
\begin{equation}
   \beta \Delta E_Z^\tau = 4 \left(K_a + K_{wlc} \right) (1-\cos \psi).
    \label{def:DEZtau}
\end{equation}
Note that for small angles ($|\phi|, |\psi| \ll 1$) the deformation
energies \eqref{def:DEVrho}, \eqref{def:DECtau}, \eqref{def:DEZtau}
are quadratic in $\phi$ and $\psi$ to lowest order.

Although the above estimates are very simple, they already illustrate
some basic features of the model.  Roll degrees of freedom are expected
to behave as the standard ``local'' WLC model. As in that case neighboring
rolls involving two consecutive angles say $\phi_n$ and $\phi_{n+1}$ will
be independent: their contribution to the total energy of the molecule
is expected to be additive. The situation is very different for tilt.
Especially in the case of very large stretching or stacking stiffnesses
(large $K_s$ or $K_{st}$), the system will avoid energetically costly
C-bends and favor a large number of Z-bends. Consecutive tilt angles
$\psi_n$ and $\psi_{n+1}$ will tend to have opposite signs (they are
anti-correlated). This induces an effective coupling between them.
As a result of the proliferation of Z-bends the ladder will be soft at
short scales and stiffer at long scales. Summarizing, the calculations of
these three elementary excitations indicate that the roll stiffness will
be very weakly dependent on $q$, while the tilt stiffness is expected
to have a maximum at $q=0$ (long wavelength behavior). This is indeed
in line with the roll and tilt behavior of DNA and RNA reported in
Fig.~\ref{fig:DNA-RNA}(a,b).

\subsubsection{Triad definition}

In order to calculate stiffness matrices from simulations we need to
define bending and twist angles.  This is done first by associating
local orthonormal triads to each pair of consecutive rungs. This is
the procedure usually followed in all-atom \cite{lave09} as well as in
coarse-grained models \cite{skor17}.  For the ladder model we define a
local tangent vector as follows
\begin{equation}
  \hat{e}_{3,n}\equiv\frac{\hat{u}_n+\hat{v}_n}{|\hat{u}_n+\hat{v}_n|}   
\end{equation}
In the above the unit vectors $\hat{u}_n$ and $\hat{v}_n$ describe the
local orientation of the individual backbones. The second component is
defined through the normal of the average plane formed by two subsequent
rungs:
\begin{equation}
    \Vec{n}_n\equiv 
    \frac{\Vec{x}_{n}\times\Vec{v}_n}{|\Vec{x}_{n}\times\Vec{v}_n|}
    +\frac{\Vec{x}_{n}\times\Vec{u}_n}{|\Vec{x}_{n}\times\Vec{u}_n|}
\end{equation}
Such that: 
\begin{equation}
    \hat{e}_{1,n} \equiv 
    \frac{\Vec{n}_n-(\Vec{n}_n\cdot\hat{e}_{3,n})\hat{e}_{3,n}}
	{|\Vec{n}_n-(\Vec{n}_n\cdot\hat{e}_{3,n})\hat{e}_{3,n}|}
    \label{eq:triad1_e1}
\end{equation}
is a unit vector normal to $\hat{e}_{3,n}$.
The last component of the triad can be defined from the prior elements
through a vector product:
\begin{equation}
    \hat{e}_{2,n} \equiv \hat{e}_{3,n} \times \hat{e}_{1,n}
\end{equation}
By construction ${\mathcal
T}_n\equiv\{\hat{e}_{1,n},\hat{e}_{2,n},\hat{e}_{3,n}\}$ defines
a local orthonormal triad. Consecutive triads $\mathcal{T}_n$ and
$\mathcal{T}_{n+1}$ can be mapped onto one another via a rotation
characterized by an Euler vector $\vec\Theta_n$. The direction
of $\vec\Theta_n$ defines the rotation axis and its magnitude
$|\vec\Theta_n|$ is the rotation angle, following the right hand rule.
The components of $\vec\Theta_n$ in terms of the basis of $\mathcal{T}_n$
define the tilt, roll and twist components
\begin{equation}
    \vec\Theta_n=l_0\left(\tau_n\hat{e}_{1,n}+\rho_n\hat{e}_{2,n}+
	\Omega_n\hat{e}_{3,n}\right)
    \label{def:deformations_ladder}
\end{equation}
We note that $\tau_n$, $\rho_n$ and $\Omega_n$ have the dimensions of
an inverse length and that $l_0\tau_n$, $l_0\rho_n$ and $l_0\Omega_n$
are the tilt, roll and twist angles expressed in radians.  The triad
definition is not unique and alternatives are possible. Different triad
definitions typically influence the definition of short scale stiffness
constants \cite{skor17}, but not the long wavelength behavior $q \to 0$.

\begin{figure}[t]
\includegraphics[width=\linewidth]{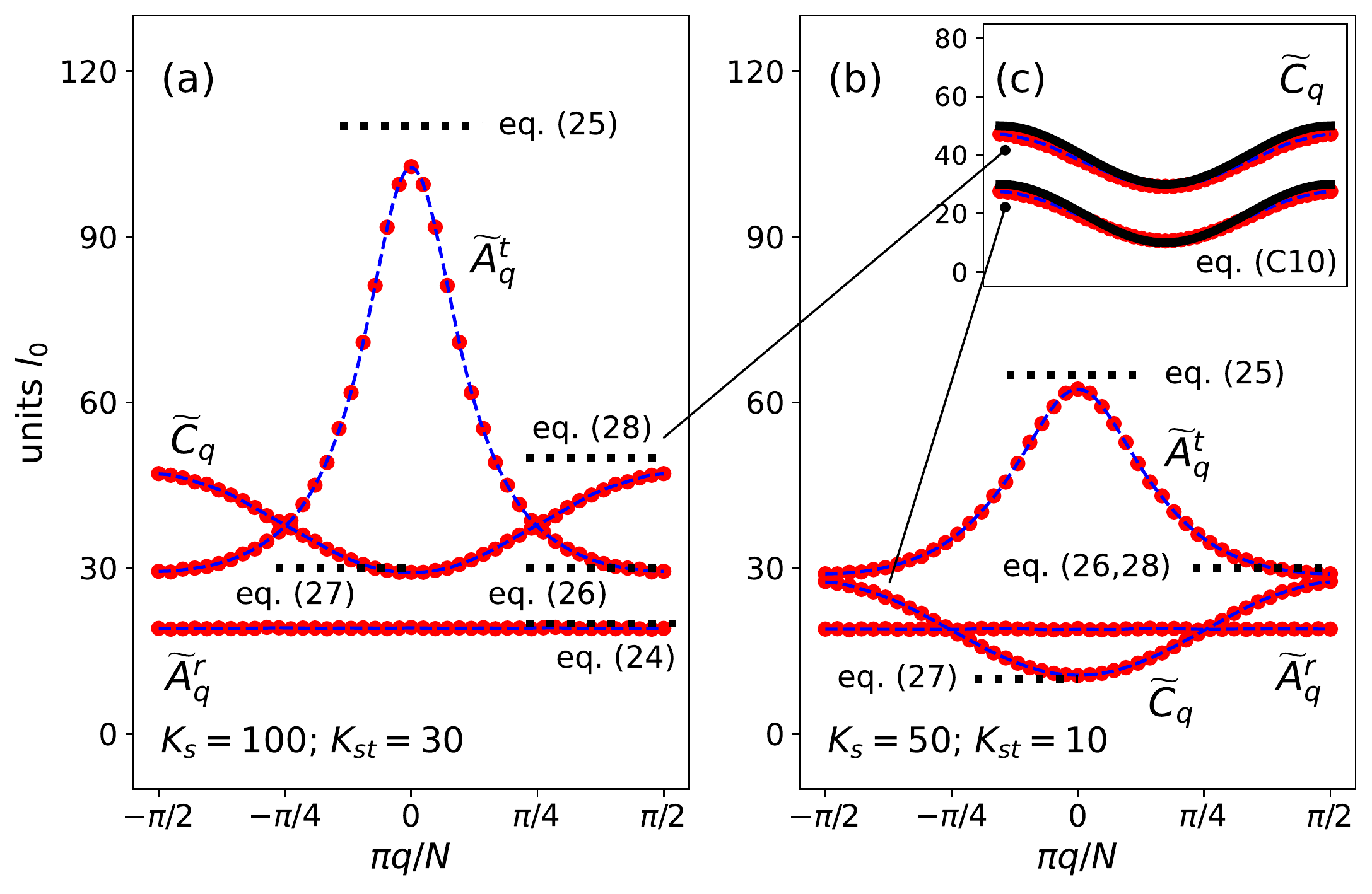}
\caption{Comparison of momentum space elasticity coefficients for the
ladder model with $K_{wlc}=K_a=10$ and (a) $K_s=100$, $K_{st}=30$ (b)
$K_s=50$, $K_{st}=10$. Both of the simulated ladders consisted of 50
rungs.  \ec{The dotted lines are the estimates of $\widetilde{A}_q^r$
from Eq.~\eqref{eq:roll}, and of at $\widetilde{A}_q^t$ and
$\widetilde{C}_q$ at $q=0$ and $q=q_\text{max}$ from \eqref{eq:tilt0},
\eqref{eq:tilt-max}, \eqref{eq:twist0} and \eqref{eq:twist-max}. The
inset (c) compares the q-dependence of the twist stiffness derived
for a general twist q-mode (see Appendix \ref{appendix:twist})
given by Eq.~\eqref{appC:Cq_calculation} (black) with the results
obtained from both simulations.} }
\label{fig:local_ladder}
\end{figure}

\subsubsection{Stiffnesses from Monte Carlo simulations}

We performed Monte Carlo simulations of the ladder model
\eqref{eq:total_energy_ladder_model}.  In these simulations a new
configuration is generated using a combination of local and global
moves which displace masses, or change local bending and twist
angles. Configurations are accepted/rejected using the Metropolis
algorithm.  Tilt, roll and twist were calculated from equilibrium
sampling using \eqref{def:deformations_ladder}.  From discrete Fourier
transformation we obtained the $q$-stiffness matrix $\widetilde{M}_q$
using \eqref{eq:covariance}.

Figure~\ref{fig:local_ladder} shows a plot of the diagonal elements
of $\widetilde{M}_q$ for two different sets of parameters. We find
virtually no off-diagonal components in the stiffness matrix of the
ladder model, so our discussion from now on will be restricted to
diagonal terms. We note that the roll $q$-stiffness $\widetilde{A}_q^r$
in Fig.~\ref{fig:local_ladder} is basically independent on $q$. This
is a consequence of the local and additive nature of roll deformations
discussed in the previous section. Since the roll is essentially due to
independent V-bends as that in Fig.~\ref{fig:ladder_deformations}(a),
the Fourier transform of Eq.~\eqref{def:DEVrho}, at small angles, gives
a $q$-independent roll stiffness
\begin{equation}
    \widetilde{A}_q^r = 2 K_\text{wlc}
    \label{eq:roll}
\end{equation}
which reproduces well the data of Fig.~\ref{fig:local_ladder}.

\begin{figure}[t]
\includegraphics[width=\linewidth]{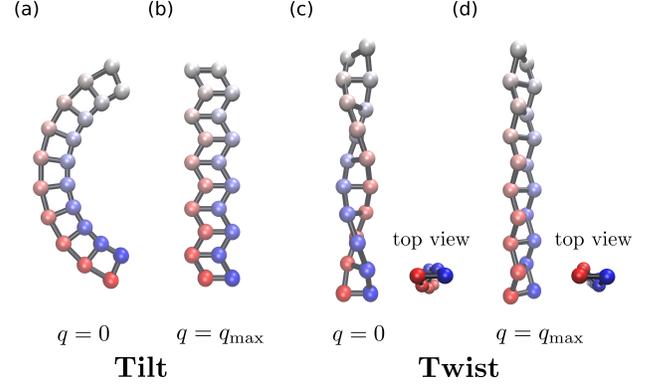}
\caption{Ladder configurations corresponding to distinct $q$-modes:
$q=0$ for (a,c) and $q=(N-1)/2$ for (b,d). (a,b) and (c,d) represent
pure tilt and twist deformations, respectively. The ladders displayed
are small subsystems extracted from infinitely long ladders, such that
on the scale of the figure angular modulation for (b,d) are omitted.}
\label{fig:Q-space_Deformations}
\end{figure}

The tilt stiffness $\widetilde{A}_q^t$ instead exhibits a marked
dependence on $q$, with a maximum at $q=0$.  The comparison between
Fig.~\ref{fig:local_ladder}(a) and (b) shows that the $q$-dependence
becomes stronger as the stretching and stacking couplings ($K_s$ and
$K_{st}$) increase. To understand the tilt behavior we estimate the
energy associated to the modes $q=0$ and $q=q_\text{max} = (N-1)/2$. Let
us consider first a pure $q=0$ mode, i.e.\ an excitation of the type
$\widetilde{\tau}_q = N \alpha \, \delta_{q,0}$, with $\delta_{m,n}$
the Kronecker delta and $\alpha$ an amplitude\ec{, where without loss
of generality $\alpha$ can be assumed to be real}.  Inverse Fourier
transforming gives ${\tau}_n = \alpha$, corresponding to a deformation
with constant tilt angle $l_0 \alpha$ at every site. This deformation is
shown in Fig.~\ref{fig:Q-space_Deformations}(a).  We can estimate its
energy as done for the C-bend deformation \eqref{def:DECtau}, assuming
that such configuration is obtained as a sequence of symmetric trapezoids
as that of Fig.~\ref{fig:ladder_deformations}(b) in the limit of small
angles. From this, we identify the stiffness of this mode
\begin{equation}
    \widetilde{A}^t_{q=0} = \frac{K_s l_0^2}{2} + 2 K_{wlc} + K_a + K_{st} 
    \label{eq:tilt0}
\end{equation}
In the following we consider a deformation of the type $\widetilde{\tau}_q
= \frac{N}{2} \left[\alpha \delta_{q,q_\text{max}}+\alpha
\delta_{q,-q_\text{max}}\right]$.  The inverse Fourier transform of the
$q_{\text{max}}$-mode reads $\tau_n=(-1)^n\alpha\cos\left(\frac{\pi
n}{N}\right)$, which corresponds to a sequence of angles having
alternating signs and a magnitude $\alpha$ which is modulated with
$\cos\left(\frac{\pi n}{N}\right)$.

Figure \ref{fig:Q-space_Deformations}(b) displays a short segment of
an infinitely long ladder compliant with the $q_{\text{max}}$-mode.
As $\cos\left(\frac{\pi n}{N}\right)$ is a weakly varying function for
large $N$, it is rendered constant on the length-scale depicted in the
figure.  We assume this configuration is obtained by a sequence of Z-bends
of Fig.~\ref{fig:ladder_deformations}(c).  We note that if $l_0 \alpha$
is the angle between consecutive backbone vectors, the angles within
each square are $\theta_{i,n} = \pi/2 \pm l_0 \alpha/2$. Following the
same calculation as \eqref{def:DEZtau} we find the following stiffness
for the mode $q_\text{max}$
\begin{equation}
    \widetilde{A}^t_{q=q_\text{max}} = 2 K_{wlc} + K_a
    \label{eq:tilt-max}
\end{equation}
Equations \eqref{eq:tilt0} and \eqref{eq:tilt-max} approximate well the
maximum ($q=0$) and the minimum ($q=q_\text{max}$) of the tilt stiffness,
as shown in Fig.~\ref{fig:local_ladder}. The slight overestimation of
Eq.~(\ref{eq:tilt0}) may be ascribed to the entropy effect, which is
neglected in the above estimation.

There is a striking difference in the behavior of $\widetilde{C}_q$
between nucleic acids and ladder model.  In DNA and RNA
the twist stiffness has a maximum at $q=0$, indicating that
these molecules are torsionally stiffer at long distances, see
Fig.~\ref{fig:DNA-RNA}(a,b) and Fig.~\ref{fig:Persistence_lengths}. The
ladder model \eqref{eq:total_energy_ladder_model} is torsionally
softer at long distances, $\widetilde{C}_q$ being minimal in the
long wavelength limit $q \to 0$, see Fig.~\ref{fig:local_ladder}. To
understand this behavior we have calculated the twist stiffness
for the mode $q=0$ and $q=q_\text{max}$, which are shown in
Fig.~\ref{fig:Q-space_Deformations}(c) and (d), respectively. As seen
for the tilt, the $q=0$ mode has a constant twist angle, while the
$q=q_\text{max}$ is formed by a sequence of twist angles of magnitude
weighed by $\cos{\left(\frac{\pi n}{N}\right)}$ and alternating signs. We
find (for details see Appendix \ref{appendix:twist})
\begin{equation}
    \widetilde{C}_{q=0}=K_{st}
    \label{eq:twist0}
\end{equation}
and
\begin{equation}
    \widetilde{C}_{q=q_\text{max}}=2K_{wlc}+K_{st}
    \label{eq:twist-max}
\end{equation}
The two previous equations are in very good agreement with the data
of Fig.~\ref{fig:local_ladder}.  Compared to the $q=0$ case, the
ladder backbone is more strongly bent in the mode $q=q_\text{max}$,
which underlies the observed property $\widetilde{C}_{q=0} <
\widetilde{C}_{q=q_\text{max}}$.  The backbone bending at $q=0$
contributes only with higher order anharmonic terms to the
energy (see Appendix~\ref{appendix:twist}), hence the stiffness
$\widetilde{C}_{q=q_\text{max}}$ is independent of the backbone
rigidity $K_{wlc}$. \ec{For the twist stiffness, it is possible
to generalize the calculations to all $q$. The estimate is given in
Appendix \ref{appendix:twist}.  A comparison with the simulations is
shown in Fig.~\ref{fig:local_ladder}(c).  }

\begin{figure}[t]
\includegraphics[width=\linewidth]{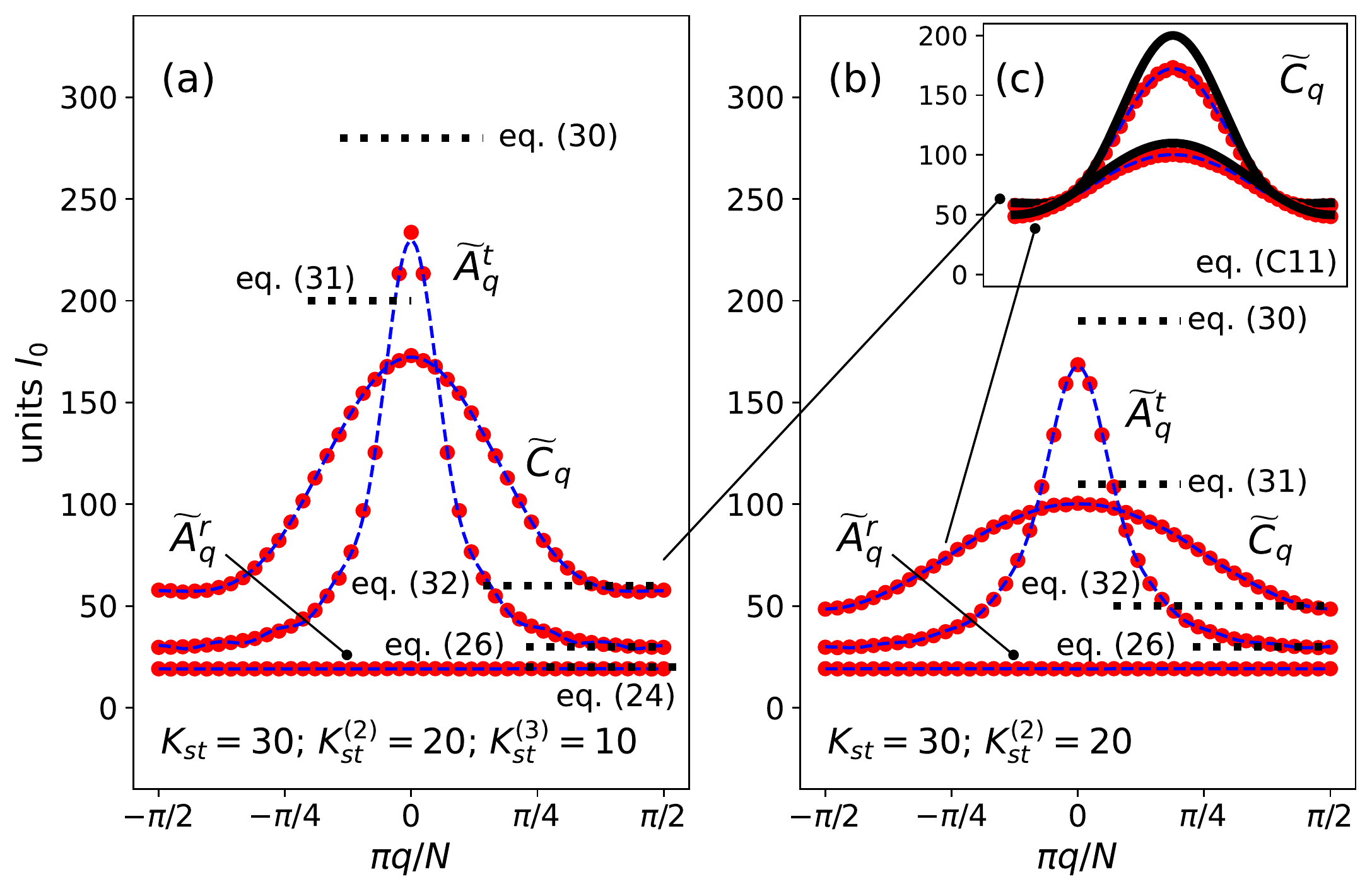}
\caption{Comparison of momentum space elasticity coefficients of two
non-local ladders having a different degree of non-locality. \ec{(a)
Stacking interactions range between nearest, next-nearest and
next-next-nearest neighbors with corresponding magnitudes $K_{st}=30$,
$K_{st}^{(2)}=20$ and $K_{st}^{(3)}=10$. (b) Stacking interactions
between nearest and next-nearest neighbors with strengths $K_{st}=30$
and $K_{st}^{(2)}=20$ respectively. } Other microscopic stiffnesses were
identical for both ladders: $K_{wlc}=K_a=10$ and $K_s=100$. Each of the
simulated ladders was comprised of 50 rungs.  \ec{The dotted lines are
the estimates of $\widetilde{A}_q^r$ from Eq.~\eqref{eq:roll}, and of at
$\widetilde{A}_q^t$ and $\widetilde{C}_q$ at $q=0$ and $q=q_\text{max}$
from \eqref{eq:tilt0-nl}, \eqref{eq:tilt-max}, \eqref{eq:twist0-nl}
and \eqref{eq:twist-max-nl}. The inset (c) shows a comparison of Monte
Carlo simulations data for the twist stiffness (red circles) with
Eq.~\eqref{appC:Cq_calculation_nl} (thick black line).  } }
\label{fig:non-local_ladder}
\end{figure}

Having explained the origin of the $q=0$ minimum of the
twist stiffness, one may ask if it is possible to extend the
model \eqref{eq:total_energy_ladder_model} to have a maximum in
$\widetilde{C}_q$ at $q=0$, as seen in nucleic acids data. To achieve
this we consider a model where stacking interactions are extended to
further neighbors:
\begin{equation}
    \beta E^* = \beta E - \sum_n \sum_{l>1} K_{st}^{(l)} \widehat{x}_n \cdot \widehat{x}_{n+l} 
    \label{eq:non-local}
\end{equation}
where $\beta E$ is the energy of the model
\eqref{eq:total_energy_ladder_model}. In practice at most two
additional couplings $K_{st}^{(2)}$ and $K_{st}^{(3)}$ were considered.
A $K_{st}^{(2)} > 0$ favors alignment of next-neighboring rungs
$\vec{x}_n$ and $\vec{x}_{n+2}$, thereby penalizing the tilt and twist
modes $q=0$. A simple calculation shows that
\begin{equation}
    \widetilde{A}^t_{q=0} = \frac{K_s l_0^2}{2} + 2 K_{wlc} + K_a + K_{st} + \sum_{l>1} l^2 K_{st}^{(l)}
    \label{eq:tilt0-nl}
\end{equation}
and
\begin{equation}
    \widetilde{C}_{q=0}=K_{st} + \sum_{l>1} l^2 K_{st}^{(l)}
    \label{eq:twist0-nl}
\end{equation}
For the $q_\text{max}$ mode we have instead
\begin{equation}
    \widetilde{C}_{q=q_\text{max}}=2K_{wlc}+K_{st} + \sum_{l>1} c_l K_{st}^{(l)}
    \label{eq:twist-max-nl}
\end{equation}
with $c_l=0$ for $l$ even and $c_l=1$ for $l$ odd. 
The tilt stiffness at $q=q_\text{max}$ \eqref{eq:tilt-max}
does not get affected by the additional stacking terms, as
the rungs remain parallel to each other for this deformation.
Figure~\ref{fig:non-local_ladder}\ec{(a,b)} shows the results of Monte
Carlo simulations of the model \eqref{eq:non-local} for two different
sets of parameters.  \ec{Figure~\ref{fig:non-local_ladder}\ec{(c)} shows
a detail of the comparison of the estimated twist stiffness for all $q$
(\eqref{appC:Cq_calculation_nl}, Appendix \ref{appendix:twist}).} This
minimal ladder model reproduces the main features of the $q$-stiffnesses
obtained from all-atom data of Fig.~\ref{fig:DNA-RNA}. The estimates of
the stiffnesses from Eqs.~\eqref{eq:tilt0-nl} and \eqref{eq:twist0-nl}
are in worse agreement with the data when compared with the counterparts
\eqref{eq:tilt0} and \eqref{eq:twist0} for the original model
\eqref{eq:total_energy_ladder_model}. Again, the analytical estimates
neglect entropic contributions which become more relevant when distal
interactions are considered.  However, for the parameter sets considered
the error is of about $15\%$. Although we have limited our discussion
here to a ``flat'' ladder, very similar stiffnesses as those shown in
Fig.~\ref{fig:local_ladder} and \ref{fig:non-local_ladder} are found in
a ladder model with intrinsic twist \cite{sege21a}, which mimicks the
double helical structure of nucleic acids.

\begin{figure}
\includegraphics[width=\linewidth]{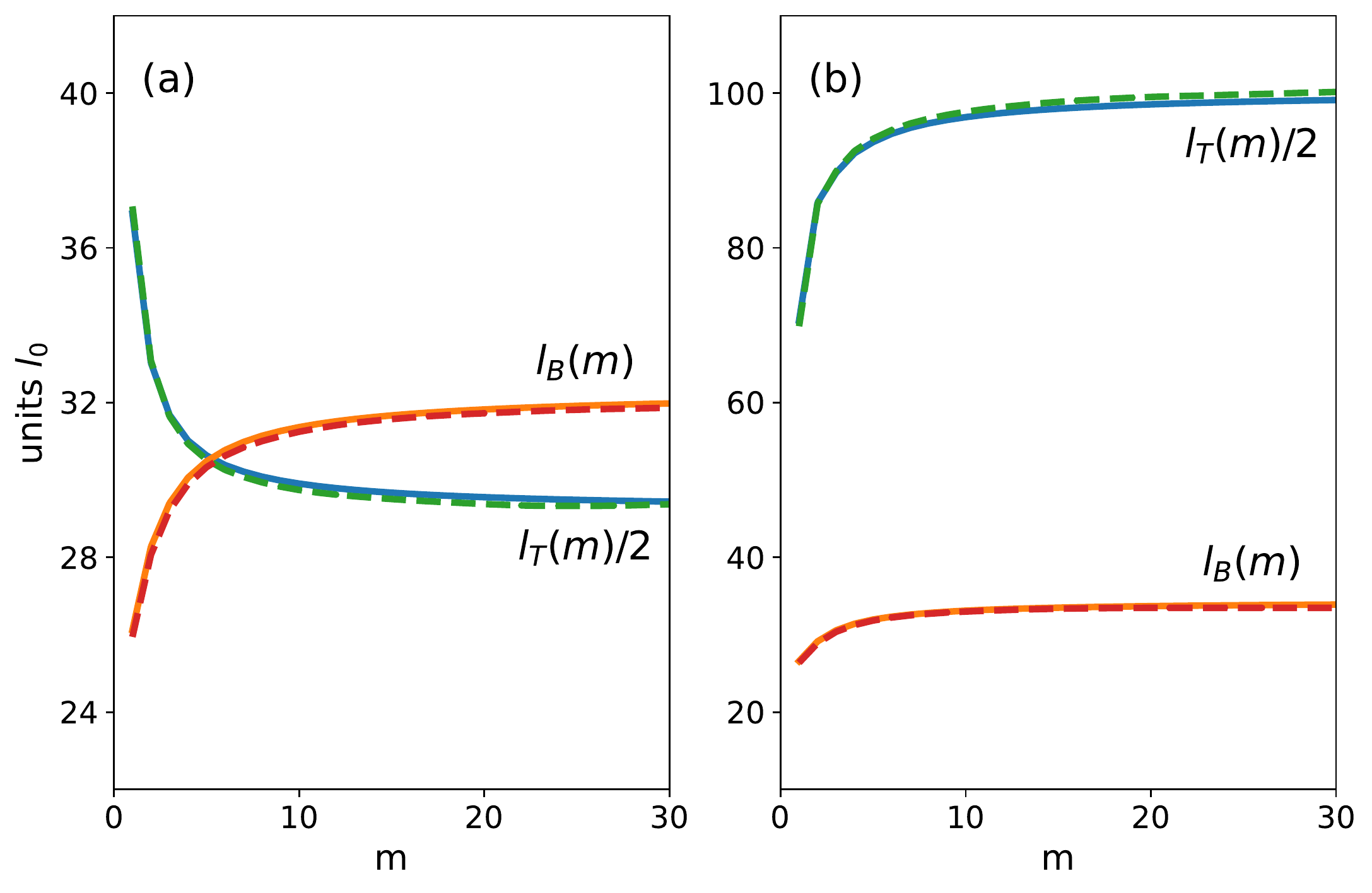}
\caption{Length-scale-dependent twist- and bending persistence
lengths attained using two distinct methods: one based on the obtained
$q$-stiffnesses using equations \eqref{RNA:lT} and \eqref{def:lBm} (solid
lines) and the other using the twist- and tangent-tangent correlation
functions defined by \eqref{eq:twist_cor} and \eqref{eq:tan-tan_cor}
(dashed lines).  The two cases correspond to ladders encompassing 50 rungs
with stiffnesses (a) $K_s = 100$, $K_{wlc} = 10$, $K_a = 10$ and $K_{st}
= 30 $ (\ec{same as Fig.~\ref{fig:local_ladder}(a)}) and (b) $K_s =
100$, $K_{wlc} = 10$, $K_a = 10$, $K_{st} = 30$ and $K_{st}^{(2)}$=20
(\ec{same as Fig.~\ref{fig:non-local_ladder}(b)}). We note that in the
case (a) the ladder is torsionally stiffer at short distances and softer
at long distances. The opposite is true in the case (b).}
\label{fig:LadderModel_PersistenceLengths}
\end{figure}

\subsubsection{Persistence lengths}

Similar to nucleic acids, the elasticity of the ladder model can be characterised 
by the bending- and twist persistence lengths. Figure \ref{fig:LadderModel_PersistenceLengths}(a,b) 
displays the persistence lengths for a local and non-local ladder respectively as 
derived from two independent methods. The first approach utilizes the attained $q$-stiffnesses 
(\ec{Fig. \ref{fig:local_ladder}(a) and \ref{fig:non-local_ladder}(b)}) from which the twist 
and bending persistence lengths can be extracted through application of equations 
\eqref{RNA:lT} (using $\widetilde{C}_q$ instead of its helical equivalent) and \eqref{def:lBm} 
correspondingly. The obtained results are indicated with the solid lines. The second 
method consisted of the inversion of the twist and tangent-tangent correlation functions given by 
formulas \eqref{eq:twist_cor} and \eqref{eq:tan-tan_cor}. Results acquired in this fashion are 
represented by the dashed lines. The good agreement between the two approaches brings additional 
support to the calculations of the RNA bending and torsional persistence lengths, where only the 
first method could be applied, as sequences analyzed are too short (Table~\ref{tab:all_all_atom_data_data}) 
to obtain the asymptotic decay of correlation functions.

\section{Discussion}
\label{sec:conclusion}

In this paper we have analyzed mechanical interactions of nucleic acids focusing on couplings
between distal sites. We have discussed these interactions in the context of the nlTWLC using 
tilt, roll and twist degrees of freedom. Non-locality in real space couplings is most conveniently
analyzed in Fourier space: translational invariance, obtained when averaging over different
sequence compositions, implies that modes with different $q$ are uncoupled. 
Using all-atom simulations we have computed the RNA $q$-stiffnesses  and compared with those of DNA, 
analyzed in a previous publication \cite{skor21}. 
\ec{Despite their close chemical structure, DNA and RNA form different helices, as illustrated 
in Fig.~\ref{fig:intro-DNA-RNA}a.} There are some strong 
\ec{qualitative} similarities \ec{in the mechanical properties of} 
DNA and RNA which can be summarized in three main points: 1) the roll stiffness is weakly dependent on $q$, 
2) tilt and twist stiffnesses are strongly $q$-dependent and 3) both molecules, in the bending and in 
the twist behavior, are soft at short length-scales and stiff at long length-scales. One difference 
between DNA and RNA is that the former has a strong non-diagonal twist-roll coupling, while this is
very weak in the latter. 
Interpolating all-atom $q$-stiffness data for short molecules (20-mers) using Fourier series we can 
use analytical expressions derived in Ref.~\onlinecite{skor21} to estimate bending and torsional 
persistence length for a long \ec{RNA} molecule. The asymptotic estimates for $l_B$ and $l_T$ are in agreement
with literature values (as we have noted, in our analysis, $l_B$ is determined with respect
to an axis perpendicular to the base pair plane and not to the helical axis). 

\ec{One fundamental prediction of the nlTWLC is the length-scale-dependent elasticity, meaning 
that the stiffness of the molecule is different at different length scales
(we note that other polymer models predict such behavior \cite{ever95,heus07}). As observed
earlier for DNA \cite{skor21}, also the RNA appears to be softer at short distances and stiffer 
at longer distances. This behavior is illustrated in Fig.~\ref{fig:Persistence_lengths}b showing
length-dependence expressed in units of base pairs
of the bending and torsional persistence lengths for RNA, as obtained
from the analysis of all-atom data. These results are qualitatively very similar to those of DNA 
\cite{skor21}. The bending persistence length is weakly length-scale-dependent, but there is 
strong dependence in the torsional response. This implies that one has to be careful to deduce
torsional stiffnesses from experiments as the results are expected to depend on the probed 
distance. Magnetic Tweezers probe the torsional properties of the two ends of $\sim 10$~kbp 
molecules \cite{lipf14}, corresponding to the asymptotic long length-scale behavior. One could 
also measure the twist stiffness by using local probes, such as by fluorescent labels. From the 
emission spectrum of a label one can deduce twist fluctuations and thus the twist stiffness 
\cite{fuji90}, but this is the torsional stiffness associated to a single base pair, corresponding 
to $m=1$ in Fig.~\ref{fig:Persistence_lengths}b. In fact local probe methods provide systematically 
lower twist stiffnesses for DNA, as opposed to the tweezers data \cite{nomi17}. 
In biological systems, double stranded nucleic acids are deformed at different length scales. 
Cells may exploit the DNA and RNA local flexibility in processes that require the 
deformation of nucleic acids at short scales. For instance, the interaction between DNA and 
DNA-bending proteins may be facilitated by local DNA flexibility.
}

We have shown that a minimal ``ladder'' model of double stranded polymer reproduces several features of 
$q$-stiffnesses observed in RNA and DNA all atom simulations. Although microscopic interactions in the ladder
are almost all of local nature (except for stacking), the analysis shows that non-locality emerges naturally 
when coarse-grained variables as tilt, roll and twist are used. The ladder model explains very naturally
the weak dependence on $q$ for the roll and it is sufficiently simple so that several analytical calculations 
are possible. These calculations link the microscopic parameters which are bond stretching and angular 
stiffnesses to the coarse-grained stiffnesses for tilt, roll and twist at least for the modes $q=0$ 
and $q=q_\text{max}$.

All-atom simulations, even when restricted to short sequences, say $20-30$-mers, require a substantial 
computational effort. Many simpler coarse-grained models have been devised \cite{dans16} to provide
accurate information about DNA mechanics, but bypassing the complexity of a simulation at all-atom scale. 
Examples are the rigid base-pair model \cite{lave09}, the rigid base model \cite{gonz13} and the
more recent multi-modal model \cite{walt20}, which goes beyond the harmonic approximation. 
These models use a set of coarse-grained coordinates and sample the DNA conformation of long molecules using
Monte Carlo methods. The parametrization is based on a large set of sequence-dependent tetranucleotide
stiffnesses. It would be interesting to test how well these coarse-grained model reproduce the sequence
dependent $q$-stiffnesses of Appendix \ref{appendix:all_atom} as these quantities sample simultaneously
the short and long scale behavior.

Finally we comment on some issues with the continuous limit of the nlTWLC. Such limit is often used in
the TWLC. We discuss here the case of a 
one component system described by a single variable $x_n$, but the discussion can be easily generalized 
to more components such as tilt, roll and twist. In $q$-space the energy is given by
\begin{equation}
    \beta E = \frac{1}{2N} \sum_q \widetilde{K}_q \, | \widetilde{x}_q |^2
\end{equation}
where the variable $\tilde{x}_q$ is obtained from the discrete Fourier transform of $x_n$. 
The continuum long wavelength limit is obtained by approximating the previous expression using
an expansion of $\widetilde{K}_q$ around $q=0$. As the energy is symmetric in $q$ the expansion gives:
\begin{equation}
    \beta E \approx \int_{-\Lambda}^\Lambda \frac{dq}{2N} \left( \widetilde{K}_0  + q^2 \Gamma \right) 
    | \widetilde{x}_q |^2
    \label{eq:continuum}
\end{equation}
where we have introduced a short scale momentum cutoff $\Lambda$. 
Let us consider the case $\Gamma > 0$ first. In the calculation of several quantities (as correlation
functions), one can safely take the $\Lambda \to \infty$ limit. Modes with large $|q|$ have high energy and
have low statistical weight, that is why many quantities are cutoff independent. Transforming back into real
space and using a continuous variable $x(s)$ one gets
\begin{equation}
    \beta E = \int_0^L ds \left[ \frac{\widetilde{K}_0}{2} x^2(s) + 
    \frac{\Gamma}{2}  \left( \partial_s x(s) \right)^2\right]
\end{equation}
This is similar to the model of twist dynamics discussed in Ref.~\onlinecite{sank05}.
However, the situation is quite different for $\Gamma < 0$, which is the relevant case for nucleic acids.
All quantities become cutoff dependent because modes such that 
\begin{equation}
    |q| > q^* = \sqrt{-\widetilde{K}_0/\Gamma}
\end{equation}
have negative energy and are unstable. Therefore the cutoff plays an essential role for $\Gamma < 0$.
Generalizing to higher dimension we note that the stiffness matrix $\widetilde{M}_q$ may contain imaginary 
off-diagonal elements which are antisymmetric in $q$. These can be written as total derivatives and do
not contribute to the continuum model limit. 

In conclusion, we have shown that length-scale-dependent stiffness is a universal property
of double stranded nucleic acids, and that such behavior is also found in a minimal ladder model.
The emergence of non-locality is mainly a consequence of coarse-graining for tilt deformations, 
while the twist behaviour is more complex in origin. Both however are well described by a nlTWLC model. 
We expect that such effects should be relevant for other polymers as well.


\acknowledgments{Discussions with E. Skoruppa are gratefully acknowledged. 
T.S was supported by JSPS KAKENHI (grant numbers: JP18H05529, 21H05759)}


\appendix

\definecolor{purple}{HTML}{800080}
\definecolor{cyan}{HTML}{00FFFF}
\definecolor{darkblue}{HTML}{00008B}
\definecolor{salmon}{HTML}{FA8072}
\definecolor{darkred}{HTML}{8B0000}
\definecolor{lime}{HTML}{00FF00}
\definecolor{chocolate}{HTML}{D2691E}

\definecolor{goldenrod}{HTML}{DAA520}
\definecolor{magenta}{HTML}{FF00FF}
\definecolor{indigo}{HTML}{4B0082}
\definecolor{royalblue}{HTML}{4169E1}
\definecolor{darkslategray}{HTML}{2F4F4F}
\definecolor{olive}{HTML}{808000}
\begin{table}[t]
\centering
\begin{tabular}{|c|c|c|c|}
\hline
Color & Sequence & &  sim. time (ns) \\ \hline
\crule[blue]{0.5cm}{0.25cm} & AAAACGAGGAUCUUAUCUCG & $^{*}$ & 10 / 100 \\
\crule[red]{0.5cm}{0.25cm} & AGCUGUGCUACCUAUAGCUG && 10 / 100 \\
\crule[green]{0.5cm}{0.25cm} & GAUCGCAAGUUGAGACCACG& & 10 / 100 \\
\crule[orange]{0.5cm}{0.25cm} & GGACACGUCGGGAGGGUUUU && 10 / 100 \\
\crule[purple]{0.5cm}{0.25cm}  & GUAACCUGGCUACGAAUGGC& & 10 / 100 \\
\crule[darkblue]{0.5cm}{0.25cm} & GCAUGCAUGACUAGCAUGCA& & 10  \\
\crule[royalblue]{0.5cm}{0.25cm} & GAUGACGUACUAGCGCAGCA& & 10   \\
\crule[salmon]{0.5cm}{0.25cm} & UUAGCAUGAUCAUAACGCAA& & 10  \\
\crule[gray]{0.5cm}{0.25cm} & GUCCACAAAGUUGAUGCUAC& & 10  \\
\crule[black]{0.5cm}{0.25cm} & GUAGCUAAUGACUAUGCAUA& & 10  \\
\crule[cyan]{0.5cm}{0.25cm} & GUAGCAUGACUGUGACACGU &$^{*}$ & 10  \\
\crule[darkred]{0.5cm}{0.25cm} & GACAUCAAUGGGACAGCACC && 10  \\
\crule[lime]{0.5cm}{0.25cm} & UCCACGCAUCAAAGCAUGUC && 10  \\
\crule[chocolate]{0.5cm}{0.25cm} & UCCGCGACAAUCUACAGUGG &$^{*}$ & 10  \\
\crule[yellow]{0.5cm}{0.25cm} & UUUGAAAUUUAUGACGUGCA && 10  \\
\crule[goldenrod]{0.5cm}{0.25cm} & ACGGUGAAAAGAUUUAACCC && 10  \\
\crule[magenta]{0.5cm}{0.25cm} & GUGUAUCGAUGUGCUACCUA && 10  \\
\crule[indigo]{0.5cm}{0.25cm} & CAAACGGUAUCACCCAACUA && 10  \\
\crule[darkslategray]{0.5cm}{0.25cm} & AGCAUCACGCCGUAUCGCAA && 10  \\
\crule[olive]{0.5cm}{0.25cm} & CGCUACCUACUGUCCGCUCG && 10  \\
\hline
\end{tabular}
\caption{List of sequences and simulation duration. 
\ec{For $5$ of $20$ sequences we performed two independent runs:
one of $10$~ns and another of $100$~ns.} 
\ec{For the remaining $15$ sequences only $10$~ns runs were performed.} 
The color in the first column  
is that used in the labels of Fig.~\ref{fig:all_all_atom_simulation}.
The three sequences labeled with a "*" show very strong deviations from
the average behavior of the $q$-stiffnesses of Fig.~\ref{fig:all_all_atom_simulation}.
These deviations involve in particular $\widetilde{A}_q^t$ and $\widetilde{C}_q$.
An analysis of these two sequences show that these have a strong multimodal
behavior.
\ec{A comparison of the 
$q$-stiffnesses for the $10$~ns and $100$~ns simulations 
for the top 5 sequences are shown in Fig.~\ref{fig:comparison_Long_Short}. }
}
\label{tab:all_all_atom_data_data}
\end{table}

\section{Details of all-atom simulations}
\label{appendix:all_atom}

Twenty all-atom simulations were performed of different oligonucleotides of dsRNA, the 
sequences are listed in Table \ref{tab:all_all_atom_data_data} along with the simulation 
time. The all-atom simulations were performed using Gromacs \cite{gromacs} (version 2018.4).
The atomic structure of the dsRNA sequences in Table \ref{tab:all_all_atom_data_data} were 
obtained with the x3DNA webtool \cite{x3dna}. This structure was put in the center of a 
dodecahedral simulation box leaving $2$~nm of free space on either side of the dsRNA molecule. 
Periodic boundary conditions were defined over the simulation box and the simulation domain 
was filled with water and $150$~mM NaCl. Additional Na$^{+}$ was added as to produce a charge 
neutral system. We used the TIP3P water model \cite{jorg83}. Before starting the simulation 
run the energy of the system was minimized such that the maximum force throughout the system 
does not exceed the threshold of 1000 kJ mol$^{-1}$nm$^{-1}$. 
Interactions were derived from the OL15 nucleic acid package \cite{zgar15} which is based on 
the amber ff99 + bsc0 force field and contains the $\chi_{OL3}$ dihedral improvement for 
RNA \cite{zgar11}. \ec{This force field was chosen for its easy availability and because it 
is a well tested standard choice \cite{dans17,spon18}. Regarding force fields for 
all-atom 
simulations of RNA it should be noted that progress has been made during past 
years \cite{tan18}. However, considering the purpose of 
our simulations was 
to demonstrate the presence of length-scale-dependent elasticity and non-local 
couplings rather than precise measurement of local parameters, the established Amber 
ff99bsc0$\chi_{OL3}$ suffices.}

Non-bonded interactions were cut-off at 1 nm and long range electrostatic 
interactions were handled through the particle-mesh Ewald algorithm. After energy minimisation, 
the system was equilibrated to a temperature of $300$~K through a molecular dynamics run 
of $100$~ps. The constant temperature was enforced by using the velocity-rescaling thermostat 
\cite{buss07}. Subsequently the system was equilibrated for $100$ ps to a pressure of $1$ 
bar where the pressure was enforced with a Parrinello-Rahman barostat \cite{parr81}. After 
equilibration, molecular dynamics simulations of 10 or 100 ns  were performed at 300 K. 
The dsRNA configurations were stored every 1 ps. \ec{The DNA data 
shown in Fig.~\ref{fig:DNA-RNA}, were taken from Ref.~\onlinecite{skor21} where they 
were generated using a very similar procedure using the parmbsc1 force field 
\cite{ivan16}, which is specific to DNA.}

\begin{figure}[t]
\includegraphics[width=\linewidth]{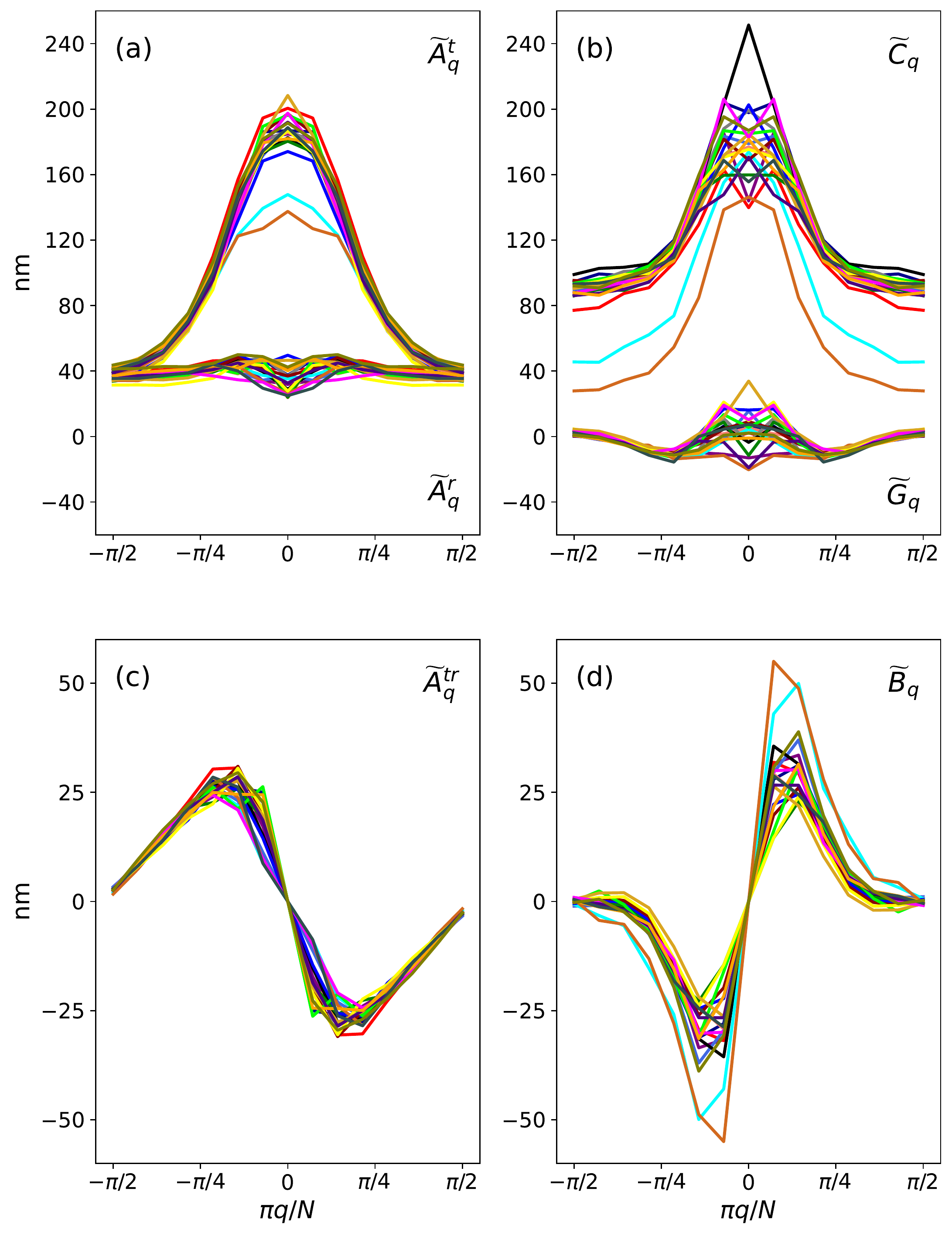}
\caption{Momentum-space stiffnesses as obtained for each individual oligonucleotide. 
The color of the plots refer to their corresponding sequence as indicated in Table 
\ref{tab:all_all_atom_data_data}. The data displayed for dsRNA in Figure \ref{fig:DNA-RNA} 
is obtained by averaging over all oligomers. }
\label{fig:all_all_atom_simulation}
\end{figure}

\begin{figure}[t]
\includegraphics[width=\linewidth]{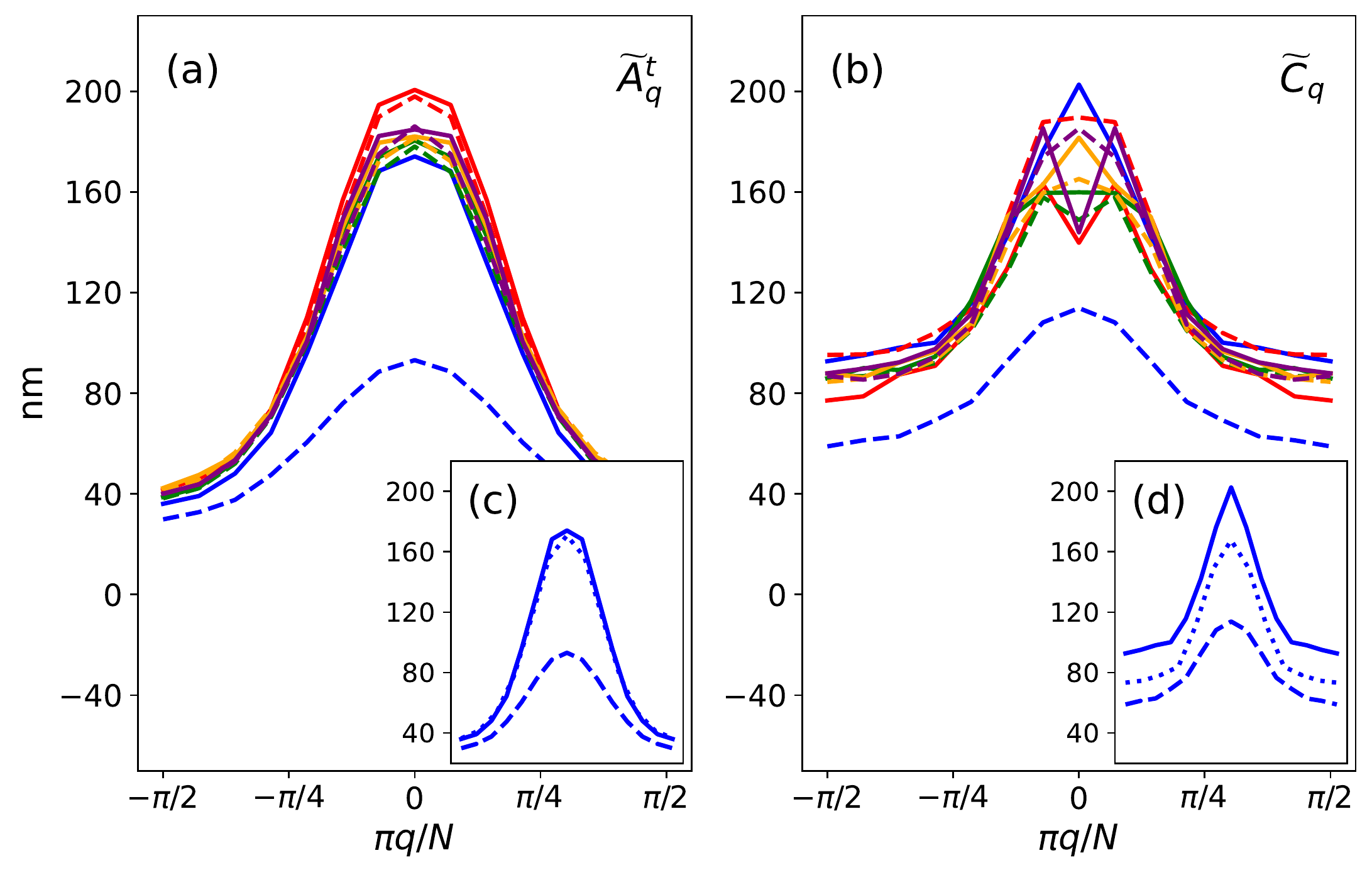}
\caption{\ec{Comparison of the momentum-space (a) tilt and (b) twist
stiffnesses obtained from 10 ns (solid) and 100 ns (dashed) all-atom
simulation for five sequences of dsRNA. The sequences are the top five of
Table \ref{tab:all_all_atom_data_data} (same color code is used).  Whilst
good correspondence between the momentum space elasticity coefficients
obtained from 10 ns and 100 ns simulations is found for four of the five
simulated sequences, substantial deviations for $\widetilde{A}^t_q$
and $\widetilde{C}_q$ are found for the AAAACGAGGAUCUUAUCUCG sequence
(blue). After inspection, the inconsistency was ascribed to a bimodality
found in the A\underline{AA}C step. The insets (c) and (d) show the
details of the stiffnesses for this sequence, showing the 10~ns simulation
data (solid) the 100~ns simulation data (dashed) and the 100~ns simulation
data with A-tract removed (dotted).  } }
\label{fig:comparison_Long_Short}
\end{figure}

The ensemble of snapshots obtained in this way, were analysed
using the Curves+ software\cite{lave09} in order to obtain an
equilibrium sampling of deformation parameters $\Delta_n$ and their
Fourier transforms from \eqref{def:Deltaq}. The momentum-space
stiffness coefficients for various oligonucleotides are shown in
Fig.~\ref{fig:all_all_atom_simulation} \ec{(for simulation times of
10~ns).  Figure~\ref{fig:comparison_Long_Short} shows a comparison between
the $10$~ns (solid lines) and the $100$~ns (dashed lines) simulation runs
for the first five sequences of Table~\ref{tab:all_all_atom_data_data}. In
four of these sequences we find good overlap between the long and
short simulation runs. One sequence (blue lines), corresponding to
the first entry in Table~\eqref{tab:all_all_atom_data_data} shows a
rather different behavior for the short and long runs, due to the
bimodal distribution associated to the A-tracts (see caption of
Fig.~\ref{fig:comparison_Long_Short} for discussion). Bimodality
\cite{dans12} here means a breakdown of the harmonic model, which
assumes gaussian fluctuations of the variables around their equilibrium
point. For long simulation runs tilt and twist show a bimodal distribution
fluctuating between two distinct values, which gives a large variance
and as a consequence low stiffness. } During analysis, the two terminal
base-pairs steps at both sides were omitted to avoid fraying effects.

\section{Real space couplings}
\label{appB}
From the stiffnesses in momentum space, the real space stiffness can be deduced by taking 
the inverse Fourier transform. Since the real-space couplings between distant sites decays 
rapidly with increasing distance, the off-site couplings can be retrieved by fitting the 
data with functions of the form:
\begin{equation}
    \Tilde{X}_q^{even}=\sum_m X_m\cos{\frac{2\pi m q}{N}}
    \label{eq:fit_even}
\end{equation}
And
\begin{equation}
    \Tilde{X}_q^{odd}=\sum_m X_m\sin{\frac{2\pi m q}{N}}
    \label{eq:fit_odd}
\end{equation}
Where $X_m$ is an element of real-space stiffness matrix $M_m$ and $\Tilde{X}_q$ the 
corresponding element in $\Tilde{M}_q$. The fits are for even and odd functions of $q$ 
respectively. The resulting real-space stiffnesses from this fitting procedure are shown 
in Table \ref{tab:RNA_real_space}. The resulting values of $A^t_m$ and $C_m$ substantiate 
the non-locality of tilt and twist deformations having significant beyond nearest-neighbour 
couplings, in contrast the range of roll-roll interactions is limited. The results 
illustrate that non-local couplings are not unique to DNA, but are also found in other 
double-helical biomolecules.

\begin{table}[t]
    \centering
\begin{tabular*}{\linewidth}{l@{\extracolsep{\fill}}cccccc}
    \hline
    \hline
    \hspace{0.1cm}N=15 & $X_0$ & $X_1$ & $X_2$ & $X_3$ & $X_4$ & $X_5$\\
    \hline
    \hspace{0.3cm}$\Tilde{A}^t_q$   & 94 & 67 & 13 & 1.0 & -1.1 & -0.5 \\
    \hspace{0.3cm}$\Tilde{A}^r_q$   & 38 & 0.9 & -1.9 & -1.5 & -0.6 & -0.2 \\
    \hspace{0.3cm}$\Tilde{C}_q$     & 112 & 41 & 15 & 2.4 & -2.1 & -2.2 \\
    \hspace{0.3cm}$\Tilde{A}^{tr}_q$& 0 & -23 & -4.4 & -1.1 & 0.3 & -0.3 \\
    \hspace{0.3cm}$\Tilde{B}_q$     & 0 & 16 & 14 & 5.7 & 1.5 & 0.3 \\
    \hspace{0.3cm}$\Tilde{G_q}$     & -4.3 & 0.3 & 6.8 & 1.1 & -0.6 & -0.4 \\
    \hspace{0.3 cm}$\Tilde{C}^{(h)}_q$ & 67 & 18 & 9.2 & 1.5 & -0.7 & -0.4 \\
    \hline
    \hline
\end{tabular*}
    \caption{Real space stiffnesses $X_m$ for dsRNA as obtained from fits of the 
    corresponding momentum space stiffnesses $\Tilde{X}_q$ displayed in 
    Figure \ref{fig:DNA-RNA}.
    }
    \label{tab:RNA_real_space}
\end{table}



\section{Twist stiffness in the ladder model}
\label{appendix:twist}

We derive here Eqs.~\eqref{eq:twist0} and \eqref{eq:twist-max} giving the twist stiffness 
of the modes $q=0$ and $q=q_\text{max}$ shown in Fig.~\ref{fig:Q-space_Deformations}(c)
and (d). Let us consider first the mode $q=0$. The positions of the masses on the two 
strands are in this case given by:
\begin{equation}
    \vec{a}^{\,\pm}_n = \frac{l_0}{2} \left[ \pm \cos(n \omega), \, \pm \sin (n \omega) , 
    \, 2 n \cos \frac{\omega}{2} \right]
    \label{appC:param-helix}
\end{equation}
These describe two helices winding around the $z$ axis and with constant twist angle $\omega$.
The diameter of the helices is $l_0$ and the pitch $\frac{2\pi l_0}{\omega} \cos (\omega/2)$.
The bond vectors (defined as in Fig.~\ref{fig:ladder_flat}) for the two strands are:
\begin{equation}
    \vec{u}_n = \vec{a}^{\,+}_{n+1}  - \vec{a}^{\,+}_n, \qquad\qquad
    \vec{v}_n = \vec{a}^{\,-}_{n+1}  - \vec{a}^{\,-}_n,
\end{equation}
while the rungs of the ladder are 
\begin{equation}
    \vec{x}_n = \vec{a}^{\,+}_{n} - \vec{a}^{\,-}_{n}
\end{equation}
The definition \eqref{appC:param-helix} implies that $|\vec{u}_n|=|\vec{v}_n|=|\vec{x}_n|=l_0$.
The stacking contribution to the energy is obtained from
\begin{equation}
    \frac{\vec{x}_n \cdot \vec{x}_{n+1}}{l_0^2} = \cos \omega \approx 1 - \frac{\omega^2}{2}
    \label{appC:stack}
\end{equation}
while to estimate the backbone bending contribution we need to compute
\begin{equation}
    \frac{\vec{u}_{n+1} \cdot \vec{u}_n}{l_0^2} = 
    1 - \left( 1 - \cos \omega \right) \sin^2 \left(\frac{\omega}{2} \right) 
    \approx 1 - \frac{\omega^4}{8}
    \label{appC:wlc}
\end{equation}
For the angles we have
\begin{equation}
    \sin \theta_{n,1} = \sqrt{1 - \left( \frac{\vec{x}_n \cdot \vec{u}_n}{l_0^2} \right)^2}
    \approx 1 - \frac{\omega^4}{32}
    \label{appC:angles}
\end{equation}
and similar relations for $\theta_{n,2}$, $\theta_{n,3}$ and $\theta_{n,4}$. 
Equations~\eqref{appC:wlc} and \eqref{appC:angles} imply that the energy of the $q=0$ 
twist mode is independent on $K_\text{wlc}$ and $K_a$, as these couple to higher order
anharmonic terms ($\sim \omega^4$). The only contribution to order $\omega^2$ comes 
from the stacking \eqref{appC:stack}, which leads to Eq.~\eqref{eq:twist0}.

The positions of the masses in the mode $q=q_\text{max}$ are
\begin{equation}
    \vec{a}^{\,\pm}_n = \frac{l_0}{2} \left[ \pm \cos \frac{\omega}{2}, 
    \pm \sin \frac{(-1)^n\omega}{2}, 2 n \cos \frac{\omega}{2} \right]
    \label{appC:param-helix2}
\end{equation}
In this case a step from site $n$ (even) to site $n+1$ (odd) corresponds to a
twist deformation with twist angle $-\omega$, while a step from an odd to an even 
site corresponds to a twist angle $+\omega$. The calculation of the stacking term 
is as above and leads to \eqref{appC:stack}. The bending of the backbone is 
different from the $q=0$ case and it is given by
\begin{equation}
    \frac{\vec{u}_{n+1} \cdot \vec{u}_n}{l_0^2} = 
    \cos \omega \approx 1 - \frac{\omega^2}{2}
    \label{appC:wlc2}
\end{equation}
We find for the angles
\begin{equation}
    \frac{\vec{x}_n \cdot \vec{u}_n}{l_0^2} = - \sin^2 \left( \frac{\omega}{2} \right) \approx
    - \frac{\omega^2}{4}
\end{equation}
which as is the case of \eqref{appC:angles} gives rise to a higher order anharmonic 
term $\sim\omega^4$ for all angular interactions. Summarizing, the harmonic contributions
to the $q_\text{max}$ mode are stacking \eqref{appC:stack} and bending \eqref{appC:wlc2}.
Combining these two one gets Eq.~\eqref{eq:twist-max}.\\
\\
\ec{In a similar fashion the calculation for the twist stiffness $\widetilde{C}_q$ can 
be extended to entail an arbitrary q-mode. For the ladder model with local stacking 
interactions such a calculation yields: 
\begin{equation}
    \widetilde{C}_q=K_{wlc} \left(1-\cos{\frac{2\pi q}{N}} \right)+K_{st}
    \label{appC:Cq_calculation}
\end{equation}
Which interpolates between the estimates of Eq.~\eqref{eq:twist0} and \eqref{eq:twist-max}. 
The estimate of Eq.~\eqref{appC:Cq_calculation} is compared with the results from Monte Carlo 
simulations in Figure \ref{fig:local_ladder}(c). For a ladder with non-local stacking 
interactions, one obtains:
\begin{eqnarray}
    \widetilde{C}_q&=&K_{wlc} \left(1-\cos{\frac{2\pi q}{N}} \right)+K_{st} 
    \nonumber \\
    && +\sum_{l>1} K_{st}^{(l)}\left(l+2\sum_{j=1}^{l-1}(l-j)\cos{\left(\frac{2\pi j q}{N}\right)}\right)
    \label{appC:Cq_calculation_nl}
\end{eqnarray}
Where the summation in $l$ runs over all non-local
stacking interactions $K_{st}^{(l)}$. For $q=0$ and
$q=q_{max}$ Eq.~\eqref{appC:Cq_calculation_nl} reduces to
\eqref{eq:twist0-nl} and \eqref{eq:twist-max-nl}, correspondingly.
In Fig.~\ref{fig:non-local_ladder}(c) the derived $q$-dependence of the
twist stiffness \eqref{appC:Cq_calculation_nl} is compared to the twist
$q$-stiffness obtained from simulations.}



%

\end{document}